\documentclass[preprint,onecolumn,nofootinbib]{revtex4}
\usepackage{amsfonts}
\usepackage{amssymb}
\usepackage{graphicx}
\usepackage{color}
\usepackage{epsfig}
\usepackage{remreset}
\usepackage{hyperref}
\usepackage{amsmath,amsthm,ulem}
\usepackage{mathtools}
\usepackage{mathrsfs}
\usepackage{color}
\usepackage{float}
\usepackage{tensor}
\usepackage{lipsum}
\usepackage[caption=false]{subfig}
\usepackage{slashed}
\usepackage{hyperref}
{
 \definecolor{BLACK}{gray}{0}
 \definecolor{WHITE}{gray}{1}
 \definecolor{RED}{rgb}{1,0,0}
 \definecolor{GREEN}{rgb}{0,1,0}
\definecolor{dgreen}{rgb}{.1,.6,.1}
\definecolor{BLUE}{rgb}{0,0,1}
 \definecolor{CYAN}{cmyk}{1,0,0,0}
 \definecolor{MAGENTA}{cmyk}{0,1,0,0}
 \definecolor{YELLOW}{cmyk}{0,0,1,0}
 \definecolor{aw}{rgb}{0.2,0.5,0.75}
 }
 \hypersetup{
  colorlinks,%
  citecolor=blue,%
  filecolor=blue,%
  linkcolor=magenta,%
  urlcolor=blue
}
\usepackage{fancyhdr}
\usepackage{remreset}
\usepackage{ulem}

\begin{document}
\baselineskip=0.5 cm

\title{Entanglement Wedge Cross-Section with Gauss-Bonnet Corrections and Thermal Quench}

\author{Yong-Zhuang Li$ ^{1}$}
%\email{liyongzhuang@just.edu.cn}
\author{Cheng-Yong Zhang$^{2}$}
%\email{ zhangcy@email.jnu.edu.cn}
\author{Xiao-Mei Kuang$ ^{3}$}
\email{xmeikuang@yzu.edu.cn (corresponding author)}

\affiliation{$^1$School of Science, Jiangsu University of Science and Technology, Zhenjiang 212003, China}
\affiliation{$^2$Department of Physics and Siyuan Laboratory,
Jinan University, Guangzhou 510632, China}
\affiliation{$^3$ Center for Gravitation and Cosmology, College of Physical Science and Technology, Yangzhou University, Yangzhou 225009, China}
%\affiliation{$^3$ School of Aeronautics and Astronautics, Shanghai Jiao Tong University, Shanghai 200240, China}
%\date{\today }

\begin{abstract}
The entanglement wedge cross section (EWCS) is numerically investigated statically and dynamically in a five-dimension AdS-Vaidya spacetime with Gauss-Bonnet (GB) corrections, focusing on two identical rectangular strips on the boundary. In the static case, EWCS increases as the GB coupling constant $\alpha$ increases and disentangles at small separation between two strips for smaller $\alpha$. For the dynamic case, such a monotonic relationship between EWCS and $\alpha$ holds but the two strips no longer disentangle monotonically as in the static case. In the early  thermal quenching stage, the disentanglement occurs at smaller $\alpha$ with larger separations. Two strips then disentangle at larger {separation} with larger $\alpha$ as time evolves. Our results indicate that the higher-order derivative corrections, like the entanglement measure in the dual boundary theory, also have nontrivial effects on the EWCS evolution.

~


~

\end{abstract}

\maketitle

\tableofcontents

\section{Introduction}
The interdisciplinary research in quantum information, condensed matter and quantum gravity has received widespread attention in the last decade. {Gauge/gravity duality \cite{Maldacena:1997re,Gubser:1998bc, Witten:1998qj} emerges as a crucial link between these fields, and is also a powerful tool for investigating various involved physical quantities, particularly in strongly coupled systems and quantum information theory.}

Entanglement is a significant concept in quantum field theory (QFT) and information theory. Entanglement entropy (EE) measures the quantum correlations
between a subsystem, $A$, and its complement, $B$, for a pure state. However,
it is challenging to compute EE using QFT techniques. Fortunately, holography simplifies this problem and it provides an elegant EE geometric duality from the gravity perspective.
Specially, for a spatial region $A$ in the boundary field theory, the geometric description of EE between $A$ and $B$ was proposed as Ryu-Takayanagi (RT) formula \cite{Ryu:2006bv,Takayanagi:2012kg,Lewkowycz:2013nqa},
%%%
\begin{equation}
S_A = \frac{\texttt{Area}(\gamma_A)}{4 G_N}\,,
\end{equation}
where $G_N$ is the bulk Newton constant of $N-$dimensional Einstein-Hilbert gravity, and $\gamma_A$ is the codimension$-2$ spacelike minimal hypersurface in the bulk geometry whose boundary is asymptotic to that of $A$. In the covariant case, the RT formula is then extended as Hubeny-Rangamani-Takayanagi (HRT) formula \cite{Hubeny:2007xt,Dong:2016hjy}.

However, EE is not a convenient quantity to measure the correlation when $A$ and $B$ are two disjoint subsystems since the total system is not pure but mixed state. Thus, a well known physical quantity to describe both the classical and quantum correlations between the subsystems $A$ and $B$ is the
mutual information (MI), which is defined as $I(A,B) = S_A + S_B - S_{A\cup B}$.
The MI is closely related to EE and is the linear combination of EE. However, MI is free from UV divergences and {the} subadditivity guarantees its nonnegativity. Moreover, in {conformal field theory (CFT),} MI
can extract more refined information than EE \cite{Calabrese:2009ez,Calabrese:2010he}. Further, it is straightforward to employ the (H)RT surfaces to study MI in a holographic framework since MI is defined in terms of EE.

More physical quantities have been constructed to measure the correlations between
the subsystems, $A$ and $B$, besides EE and MI, particularly for the mixed state, for instance, entanglement of purification ($E_P$) \cite{Terhal:eop},
reflected entropy ($S_R$) \cite{Dutta:2019gen}, odd entropy ($S_O$) \cite{Tamaoka:2018ned} and logarithmic negativity($\xi$) \cite{Plenio:2005cwa}. Their definitions in QFT are briefly reviewed in {Appendix} A.  In the holographic framework, all these correlation measures are connected with the entanglement wedge cross-section (EWCS) {via}
\begin{equation}
E_W=\frac{\texttt{Area}(\Sigma_{AB}^{\texttt{min}})}{4G_N},
\end{equation}
where $\Sigma_{AB}^{\texttt{min}}$ denotes the minimal cross sectional area of the entanglement wedge.
$E_P$ and $S_R$ are good measures of total correlations between two disjoined
subsystems. As shown in {Appendix A}, their definitions depend on the purification schemes of the mixed states. Their holographic connections with EWCS were investigated in \cite{Takayanagi:2017knl} and \cite{Dutta:2019gen} where the authors conjectured $E_P=E_W$ and $S_R=2E_W$, respectively. {$S_O$ is a new measure of total correlations for mixed states, and its holographic description is $S_O(A,B)=S(A,B)+E_W(A,B)$.} However, $\xi$  captures only the quantum correlation for mixed states, which is different from $E_P, S_R$ and $S_O$. The conjectured holographic relationship between the logarithmic negativity and EWCS was investigated in \cite{Plenio:2005cwa,Kusuki:2019zsp}, where it demonstrated that for the vacuum state and ball-shaped
subregions, $\xi=\chi_d E_W$ with $\chi_d$ constant determined by the theory's dimension. Besides, the authors of \cite{Agon:2018lwq} linked the EWCS to entanglement distillation using bit threads. {Thus, EWCS is a critical bulk geometry description of the correlations of mixed states in the dual boundary theory. More efforts of analytical or numerical studies have been conducted in this direction in \cite{Espindola:2018ozt,Yang:2018gfq,Liu:2019qje,Ghodrati:2019hnn,BabaeiVelni:2019pkw,Chakrabortty:2020ptb,
BabaeiVelni:2020wfl,Fu:2020oep,Gong:2020pse,Lala:2020lcp,Liu:2020blk,Jokela:2020wgs,Khoeini-Moghaddam:2020ymm,Sahraei:2021wqn} and therein.}

{Further, nonequilibrium dynamics and system thermalization in QFT have piqued the interest of researchers because they can be used to describe processes in various areas of physics. It is significant to have an in-depth understanding of thermalization, which is the process whereby physical quantities of the system attain equilibrium after a certain perturbation acts upon its  initial equilibrium state. Thermalization is always implemented by a quenching process in the system, and it is holographically described by the black hole formation from the gravitational collapse, which is widely employed as an effective model to study the thermalization process, see \cite{Balasubramanian:2010ce,Balasubramanian:2011ur} as a review.
Moreover, because a mixed state must remain mixed as time evolves, it is significant to explore how the final mixed state will replicate a thermal one. In {QFT}, the dynamic evolution of
reflected entropy, logarithmic negativity and odd entropy after global/local quench and their holographic dual have been analyzed in \cite{Kusuki:2019evw,Kusuki:2019rbk,Moosa:2020vcs,Kudler-Flam:2020url}. In addition, the EWCS time evolution in Vaidya geometry has been investigated in \cite{Yang:2018gfq,BabaeiVelni:2020wfl}.}

{Beyond the Einstein gravity, it is natural to generalize the aforementioned investigations into higher derivative gravity, where the higher derivative corrections in the bulk are dual to the large N or the large coupling constant corrections in the boundary field theory from AdS/CFT {correspondence} perspective \cite{Aharony1999prd,Aharony2000RP,Otero2015}. The main goal of this paper is to investigate EWCS with Gauss-Bonnet (GB) corrections in both static and dynamic cases. On the one hand, Einstein-Gauss-Bonnet gravity theory is the simplest case of higher-derivative Lovelock gravity, {and it has an exact Vaidya type black brane solution} \cite{Rong2002PRD, Dominguez:2005rt}. All other curvature square corrections in Einstein gravity can be reduced to a GB term by redefining fields and disregarding six or more derivatives \cite{Brigante2008PRL,Brigante2008PRD}. This provides the opportunity to study the dual boundary theory's holographic properties of higher-order corrections. On the other hand, a comprehensive and profound understanding of the duality between higher derivative gravity and the boundary field theory will aid us in validating the holographic duality principle. Several investigations have been conducted on this topic. For example, the effects of higher derivative corrections on holographic EE and MI have been investigated in \cite{Hung2011jhep,Boer2011jhep,Guo:2013aca,Dong:2013qoa,Haehl:2017sot,Tanhayi2018epjc}. In addition, their properties during the thermalization with GB corrections were explored with or without other additional conditions in \cite{Li:2013cja,Li:2013sia,Zeng2014jhep,Shao2015PRD,Sun2016jhep,Caceres2017jhep,Farsam2019GRG}. This study is expected to explain the correlations of mixed states in the boundary theory dual to Einstein-Gauss-Bonnet gravity theory. Practically, in this work we mainly employs the conjecture $E_P=E_W$ and investigates the properties of the entanglement measure, $E_P$, of the {dual} boundary theory. We will also investigate EWCS evolution in the equilibration process after a quantum quench implemented by GB-Vaidya theory.}

This paper is structured as follows. In section \ref{sec:review}, we review the general Einstein-Gauss-Bonnet black brane solution and its HEE computation. Then the construction of the holographic mutual information (HMI) and EWCS of the theory is represented in section \ref{sec:HMIandEoP}. In section \ref{sec:static result} and \ref{sec:Vaidya result}, we show the numerical results of EWCS and discuss the properties {under the influence of} GB corrections for static and Vaidya GB backgrounds. The last section presents our conclusion and discussion.

\section{Review of HEE in Einstein-Gauss-Bonnet theory}\label{sec:review}
For $d+1$ $(d\geq4)$ dimension Einstein-Gauss-Bonnet gravity dual to a $d-$dimensional boundary CFT, the action is given by
\begin{eqnarray}
I=\frac{1}{16 \pi G^{d+1}_{N}}\int d^{d+1}x \sqrt{-g}\,[\mathcal{R}-\Lambda+\frac{\alpha\,L^{2}}{(d-2)(d-3)}\mathcal{L}_{GB}],
\label{ActionGB}
\end{eqnarray}
where
\begin{eqnarray}
\Lambda&=&-\frac{d(d-1)}{L^{2}},\\
\mathcal{L}_{GB}&=&\mathcal{R}^{2}-4 \mathcal{R}_{\mu\nu}\mathcal{R}^{\mu\nu}+\mathcal{R}_{\mu\nu\rho\sigma}\mathcal{R}^{\mu\nu\rho\sigma}.
\end{eqnarray}
$\alpha$ is the GB coupling constant and $L$ is the AdS spacetime radius (hereinafter, $L=1$). A constraint $-\frac{(d-2)(3d+2)}{4(d+2)^2}\leq\alpha\leq \frac{(d-2)(d-3)[(d-2)^2+3d+2]}{4[(d-2)^2+d+2]^2}$ restricted by the dual boundary field theory's causality or the positivity of the energy flux in the CFT analysis \cite{Brigante2008PRL,Buchel2009jhep,Hofman2009NPB,Ge:2009eh} exists. The above action confirms a $d+1$ dimensional static black brane solution \cite{Rong2002PRD}
\begin{eqnarray}
ds^{2}=-r^{2}f(r)dt^{2}+\frac{1}{r^{2}f(r)}dr^{2}+\frac{r^{2}}{L^{2}_{AdS}}d{\mathbf{x}}^{2},
\label{BH}
\end{eqnarray}
where
\begin{eqnarray}
f(r)=\frac{1}{2\alpha} \left[1-\sqrt{1-4\alpha(1-\frac{r_{h}^{d}}{r^{d}})}\right], \quad L_{AdS}=\sqrt{\frac{1+\sqrt{1-4\alpha}}{2}}.
\end{eqnarray}
$r_{h}$ is the event horizon radius and $L_{AdS}$ is the effective AdS radius. $\mathbf{x}=(x_{1}, x_{2}, \cdots, x_{d-1})$ {correspond} to the spatial coordinates on the boundary. With the Eddington-Finkelstein coordinates, the above solution can be expressed as follows:
\begin{eqnarray}
ds^{2}=\frac{1}{z^{2}}\left[-f(z)dv^{2}-2dzdv+\frac{1}{L_{AdS}^{2}}d\mathbf{x}^{2}\right],
\label{EFsolution}
\end{eqnarray}
where
\begin{eqnarray}
f(z)&=&\frac{1}{2\alpha}\left[1-\sqrt{1-4\alpha\left(1-\frac{z^{d}}{z_{h}^{d}}\right)}\right],\\
dt&=&dv+\frac{dz}{f(z)},\quad r=\frac{1}{z}.
\label{corrdinatestrans}
\end{eqnarray}

{In Einstein Gauss-Bonnet gravity,} for a strip $A$ in the $d$-dim boundary  field theory with one dimension $x_{1}$ of length $l$ and the other $d-2$ dimensions of volume $H^{d-2}$,\footnote{We consider the very large $H$ such that the lengths of other dimensions will not affect the behavior of the entanglement entropy.} its EE should be modified as follows:
\begin{eqnarray}
S(V)=\frac{1}{4G_{N}^{d+1}}\left[\int_{\Sigma}d\mathbf{x}^{d-1}\sqrt{h}\left(1+\frac{2\,\alpha}{(d-2)(d-3)}\mathcal{R}_{\Sigma}\right)+\frac{4 \alpha}{(d-2)(d-3)}\int_{\partial \Sigma}d\mathbf{x}^{d-2}\sqrt{\sigma}\mathcal{K}\right],
\label{EEaction}
\end{eqnarray}
where $\Sigma$ indicates a minimal surface that extends into the bulk and shares the boundary with $A$, \textit{i.e.}, $\partial \Sigma=\partial A$. $V$ represents the volume of the bulk enclosed by $\Sigma$ and $A$.  $\mathcal{R}_{\Sigma}$ is the induced scalar curvature of surface $\Sigma$, and $\sigma$ is the determinant of the induced metric of the boundary $\partial\Sigma$.
{$\mathcal{K}$ is the trace of the extrinsic curvature of  $\partial\Sigma$. Readers could refer to \cite{Hung2011jhep,Boer2011jhep,Guo:2013aca,Dong:2013qoa,Haehl:2017sot} for more details.}

The induced metric of $\Sigma$ is given as follows:
\begin{eqnarray}
ds^{2}_{\Sigma}=h_{ij}dx^{i}dx^{j}=\frac{1}{z^{2}}\left[-f(z)(v')^{2}-2z'\,v'+\frac{(x')^{2}}{L^{2}_{AdS}}\right]du^{2}+\frac{1}{z^{2}\,L^{2}_{AdS}}d\mathbf{y}^{2}.
\label{inducedmetric}
\end{eqnarray}
The prime here indicates the derivative with respect to $u$ which parameterizes the minimal surface $\Sigma$. $x$ and $\mathbf{y}$ represent the dimension $x_{1}$ and other $d-2$ dimensions, respectively. The induced scalar curvature $\mathcal{R}_{\Sigma}$ is thus given by
\begin{eqnarray}
\mathcal{R}_{\Sigma}=-\frac{(d-2)\left\{(d+1)h_{uu}(u) z'(u)^{2}+z(u)(h_{uu}'(u)z'(u)-2h_{uu}(u)z''(u))\right\}}{h_{uu}(u)^{2}z(u)^{2}},
\label{Ricci}
\end{eqnarray}
while by choosing the normalized unit vector $n_{a}=\sqrt{h_{uu}(u)}\delta_{ua}$, we obtain the following:
\begin{eqnarray}
\mathcal{K}=-\frac{d-2}{\sqrt{h_{uu}(u)} z(u)} z'(u).
\label{extrinsic}
\end{eqnarray}
Finally, Eqn.~(\ref{EEaction}) can be rewritten as follows:
\begin{eqnarray}
S(V)=\frac{H^{d-2}}{4G_{N}^{d+1}}\int_{\Sigma}du\left[\frac{\sqrt{h_{uu}(u)}}{z(u)^{d-2} L_{AdS}^{d-2}}+\frac{2\alpha z'(u)^{2}}{z(u)^{d} L_{AdS}^{d-2} \sqrt{h_{uu}(u)}}\right],
\label{EEaction2}
\end{eqnarray}
where $H^{d-2}$ indicates the volume of strip on {all} other dimensions. We will consider $d=4$ case and use $x$ as the parameter, i.e., $x'(u)=1$.

{We will adopt the thermalization process to investigate the evolutionary properties of EoP via EWCS with GB corrections, which is conventionally modeled using a homogeneous \textit{falling} thin shell of null dust in the bulk. We introduce a Vaidya-type solution \cite{Dominguez:2005rt,Rong2002PRD}:
\begin{eqnarray}
ds^2=\frac{1}{z^2}\left[-f(v,z)dv^{2}-2dzdv+\frac{1}{L^{2}_{AdS}}d\mathbf{x}^{2}\right],
\label{timedependetmetric}
\end{eqnarray}
where $f(v,z)=\frac{1}{2\alpha}[1-\sqrt{1-4\alpha(1-m(v)z^{d})}]$. The mass function is
\begin{eqnarray}
m(v)=\frac{M}{2}\left[1+\text{tanh}\left(\frac{v}{v_0}\right)\right],
\label{massfunction}
\end{eqnarray}
where $M$ denotes the mass of the black brane outside the shell, \textit{i.e.}, $v>v_0$. $v_0$ represents a finite shell thickness. We will choose $M=1$ and $v_0=0.02$ for the numerical calculation.}

\section{Setup of the entanglement wedge cross-section with GB correction}\label{sec:HMIandEoP}

We consider two identical strips $A$ and $B$ {with width $\ell$ in the boundary}. The distance between two strips is designated as $C$ with width $s$. Further, HMI is defined as follows:
\begin{eqnarray}
I[A,B]=S[A]+S[B]-\text{Min}\left(S[A\cup B]\right),
\label{hmidefinition}
\end{eqnarray}

\begin{figure}[h]
\centering
 \includegraphics[width=0.7\textwidth]{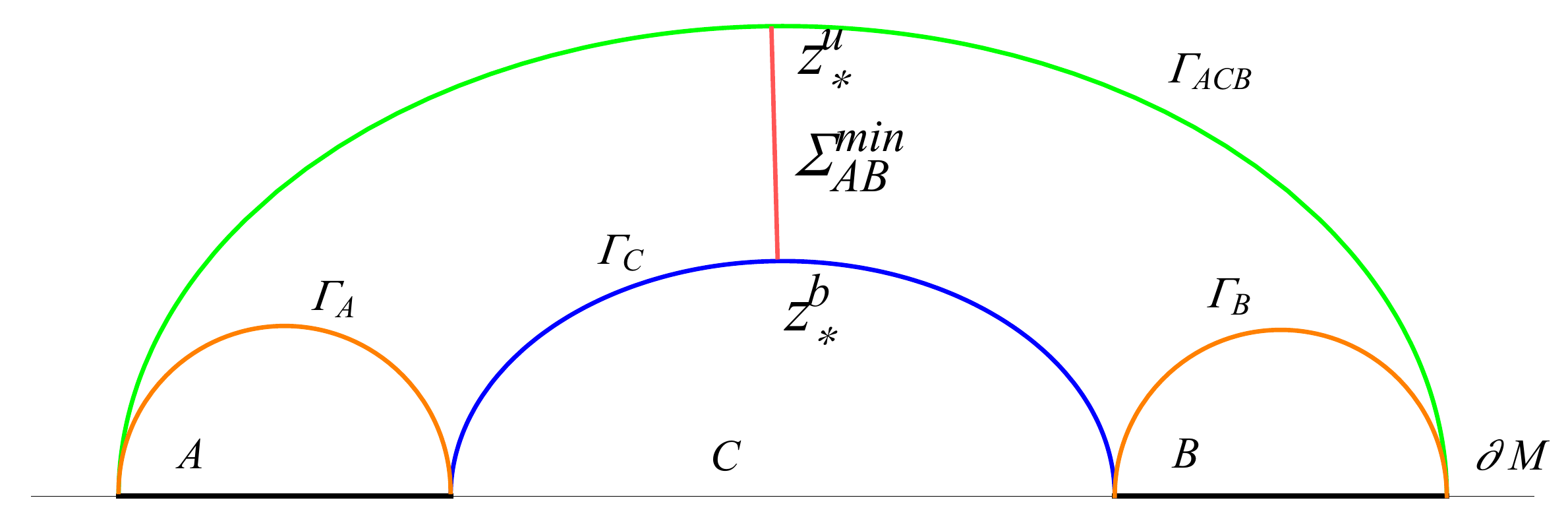}\ \hspace{0.1cm}
	  \caption{The schematic configuration for computing  {the area of EWCS (Hereafter, we will also refer the area of EWCS to EWCS, which in principle could not be confused.)} with minimal surface $\Sigma_{AB}^{min}$ when the entanglement wedge is connected. The two strips, $A$ and $B$, are identical with width $\ell$. The separation between two strips is labeled as $C$ with width $s$. $\Gamma$ indicates the extremal surface for each strip.}
 \label{fig:eop}
\end{figure}

In the following sections, we will construct the coordinate system, such that $A$ and $B$ are symmetric about the $z$-axis (Figure \ref{fig:eop}).
%%%%%

By definition, the entanglement of purification is holographically modeled by the wedge cross section in an AdS gravity \cite{Takayanagi:2017knl}. However, the impacts of the boundary term may need to be considered while investigating the wedge cross-section in an AdS-GB gravity. Consequently, there are two alternative EWCS definitions.

The first one is directly given by the standard definition, i.e., the extremal surface defined by the induced metric without the Gibbs-Hawking boundary term, which is given mathematically as follows:
\begin{eqnarray}
S_{1}(V)=\int_{m}\sqrt{h}du,
\end{eqnarray}
where integral ranges from symmetric point, $z_{*}^{b}$, on $\Gamma_C$ to point $z_{*}^{u}$ on $\Gamma_{ACB}$.

Further, the second choice is the standard definition modified by the Gibbs-Hawking term. We claim that this definition should be reasonable. Physically, the extremal surface corresponding to the EWCS should be part of an HEE extremal surface truncated by point $z_{*}^{b}$ on $\Gamma_C$ and $z_{*}^{u}$ on $\Gamma_{ACB}$. The intersection conditions are given by $z_{\perp}(p_1)\perp \Gamma_C$ and $z_{\perp}(p_2)\perp\Gamma_{ACB}$ where  $z_{\perp}(p)$ indicates the orthogonal vector of $\Gamma$ at point $p$. It is well known that the impacts of boundary Gibbs-Hawking term must be considered while investigating HEE in an AdS-Gauss-Bonnet gravity. This is the intrinsic sample property of holography, i.e., the boundary field quantities are represented by bulk geometric quantities. So the entanglement of purification with higher derivative corrections in physical intuitions should be identified by the EWCS with the boundary effects.

Since the two subsystems are identical, the minimal surface along the radial geodesic $z(0)$ from $\Gamma_C$ to $\Gamma_{ACB}$ lies in the $(z,v)$ plane. So the induced metric is given as follows
\begin{eqnarray}
ds^{2}_{eop}=\frac{1}{z^{2}}\left[-f(z,v)-2\frac{dz}{dv}\right]dv^{2}+\frac{1}{z^{2}\,L^{2}_{AdS}}d\mathbf{y}^{2}.
\label{inducedmetriceop}
\end{eqnarray}
According to Eqn.~(\ref{EEaction2}), the volume of $\Sigma_{AB}^{min}$ can be presented as
\begin{eqnarray}
\tilde{S}_{eop}(V)=\int_{\Sigma}du\left[\frac{\sqrt{h_{uu}(u)}}{z(u)^{d-2} L_{AdS}^{d-2}}+\frac{2\alpha z'(u)^{2}}{z(u)^{d} L_{AdS}^{d-2} \sqrt{h_{uu}(u)}}\right],
\label{eopaction}
\end{eqnarray}
where we have absorbed the constant coefficient $4G_{N}^{d+1}H^{2-d}$ into $\tilde{S}_{eop}(V)$. Considering Eqn. (\ref{corrdinatestrans}),  the following is given for the static scenario
\begin{eqnarray}
\tilde{S}^{st}_{eop}(V)=\int_{z_{*}^{b}}^{z_{*}^{u}}\frac{dz}{z^{d-1}L_{AdS}^{d-2}}\left[\sqrt{f(z)^{-1}}+2 \alpha \sqrt{f(z)}\right],
\label{eopactionstatic}
\end{eqnarray}
where ``$z_{*}^{u}$'' and ``$z_{*}^{b}$'' stand for $z$ at symmetric points of $\Gamma_{ACB}$ and $\Gamma_{C}$, respectively.

Further, the following is presented for the dynamic case:
\begin{eqnarray}
\tilde{S}^{dy}_{eop}(V)=\int_{v_{*}^{b}}^{v_{*}^{u}}\frac{dv}{z^{d-1}L_{AdS}^{d-2}}\left[\sqrt{-f(z,v)-2\frac{dz}{dv}}+2 \alpha \left(\frac{dz}{dv}\right)^{2} \left(\sqrt{-f(z,v)-2\frac{dz}{dv}}\right)^{-1}\right],
\label{eopactiondy}
\end{eqnarray}
where $v_{*}^{u}$ and $v_{*}^{b}$ correspond to $v$ at symmetric points of $\Gamma_{ACB}$ and $\Gamma_{C}$, respectively, and should be determined at the same boundary time. For $\alpha=0$, this equation reduces to the holographic EoP's standard representation \cite{Yang:2018gfq}.

\section{Entanglement wedge cross-section in static Gauss-Bonnet black brane}\label{sec:static result}
For static GB black brane, the re-normalized HEE is given by the following:
\begin{eqnarray}
\mathcal{S}(V)&=&\int_{\Sigma}dx\left[\frac{\Phi(x)}{z(x)^{3} L_{AdS}^{2}}+\frac{2\alpha z'(x)^{2}}{z(x)^{3} L_{AdS}^{2} \Phi(x)^{2}}\right]-\mathcal{S}_{div}(V),\\
\Phi(x)&=&\sqrt{\frac{1}{L_{AdS}^{2}}+\frac{z'(x)^{2}}{f(z(x))}},
\label{RHEE}
\end{eqnarray}
where $\mathcal{S}(V)\equiv 4 G^{d+1}_{N}S(V)/H^{d-2}$ and $\mathcal{S}_{div}(V)$ are the divergent terms due to the extremal surface divergence near the boundary.

We could obtain a complex equation of motion by extremizing this action, which will not be presented here.
The numerical method should be accepted since an analytical solution for such an equation will be impossible to discover. The initial conditions should be set in such a way that near the boundary $z\rightarrow\epsilon$ with $\epsilon$ a UV cut-off, and $z(P)\rightarrow z_{*}, z'(P)\rightarrow 0$ with $P$ indicating the symmetric point of the extremal surface. Because of the translation invariant along the $x$ direction, we only need to numerically find the relationship between the strip width and $z_{*}$ once.

%With $x$ as the parameter, the conservation equation is
%\begin{eqnarray}
%\frac{\Phi(x)^{2}-2 \,\alpha\, z'(x)^2}{L_{AdS}\, \Phi(x)^{3}}=\left(\frac{z(x)}{z_{*}}\right)^{3}.
%\label{s-conservation}
%\end{eqnarray}

\begin{figure}[h]
\centering
 \includegraphics[width=0.45\textwidth]{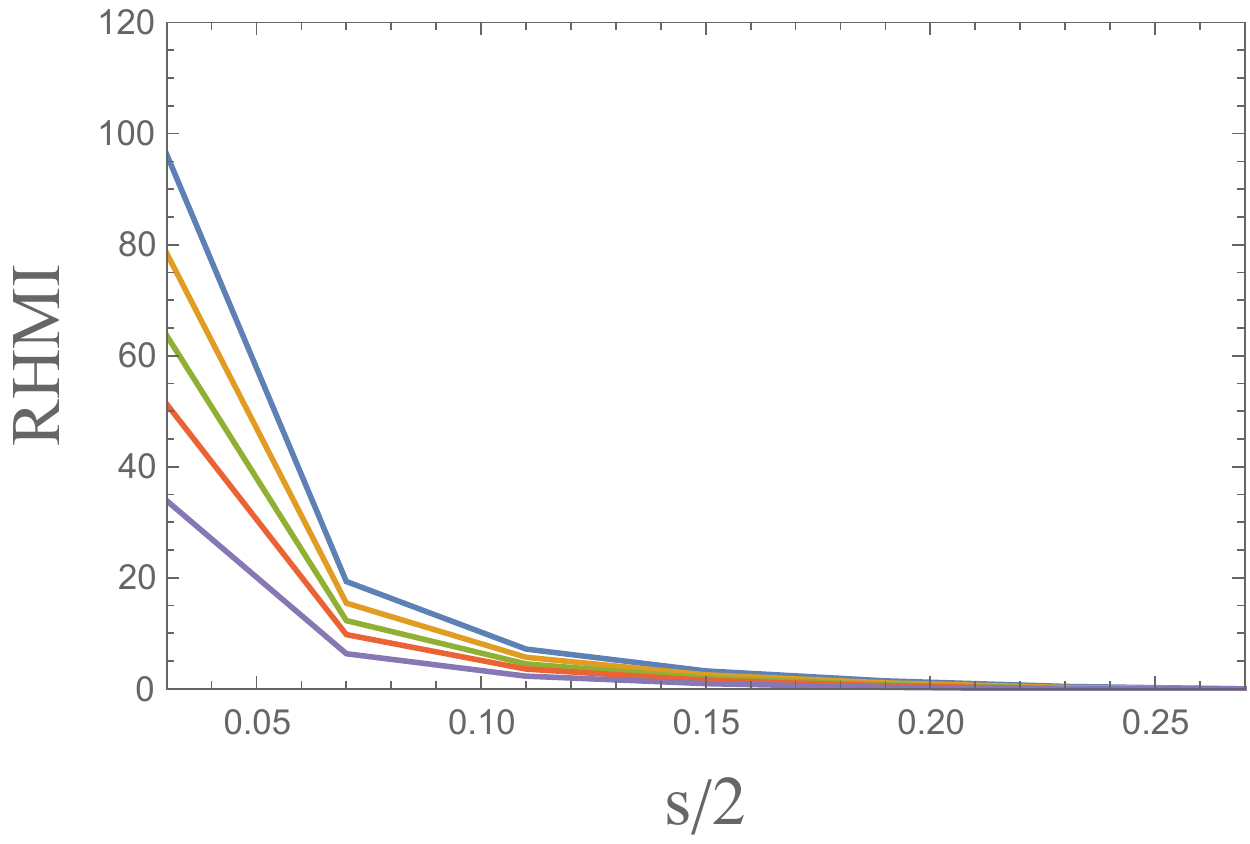}\ \hspace{0.1cm}
 \includegraphics[width=0.45\textwidth]{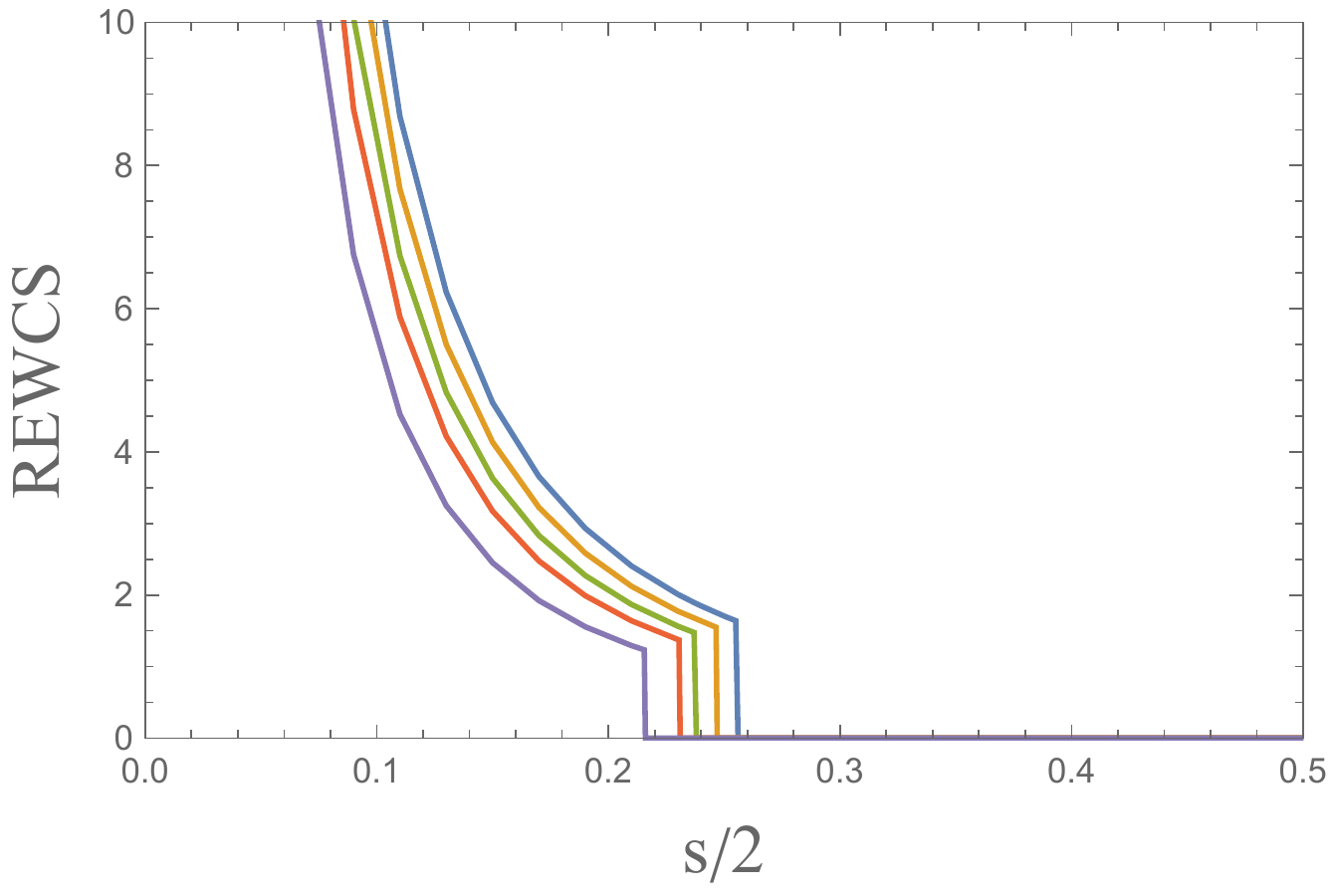}\ \hspace{0.1cm}
	  \caption{(Left) The renormalized holographic mutual information (RHMI) with the width of separation for different $\alpha$, showing a monotonic relation between RHMI and $\alpha$. From top to bottom $\alpha=0.05,0,-0.05,-0.1,-0.19$. (Right) The renormalized entanglement wedge cross-section with the distance of two strips for different $\alpha$, showing a monotonic relation between EoP and $\alpha$. From right to left $\alpha=0.05,0,-0.05,-0.1,-0.19$.}
 \label{fig:rhmi}
\end{figure}

From Figure~\ref{fig:eop}, the renormalized holographic mutual information (RHMI) is then defined by the following:
\begin{eqnarray}
I(A,B)=2 \mathcal{S}(\ell)-\text{Min}\left[2\mathcal{S}(\ell),\mathcal{S}(2\ell+s)+\mathcal{S}(s)\right].
\label{RHMI}
\end{eqnarray}
To perform the numerical calculation, we set $2\ell+s=4.98 and z_{0}=v_{0}=0.05$. In Figure~\ref{fig:rhmi} we plotted RHMI and renormalized entanglement wedge cross-section (REWCS) with different $\alpha$. For both cases, a monotonic relation reveals that with larger $\alpha$, the corresponding RHMI or REWCS is larger. {Both RHMI and REWCS go from nonzero to zero at certain strip separations ($s_c$), and for $s<s_c$, RHMI and REWCS are positive, whereas they are zero for $s>s_c$. The behavior from nonzero to zero in RHMI or REWCS could be considered a phase transition from an entangled phase to a disentangled phase occurring at $s_c$. The disconnection between $A$ and $B$ or the vanishing RHMI results in the discontinuous decrease of REWCS to zero. Notably, due to the quantum correlations between two disjoint regions, the quantum MI {between} them does not vanish, even when they are far apart \cite{Michael2008prl}. This implies that these phase transitions could be eliminated by considering the whole quantum effects{. Nevertheless,} the question of how to describe their quantum correction in the dual bulk remains unsolved, albeit a recent attempt in this regard is presented in \cite{Engelhardt2015jhep}. Moreover, it was addressed in \cite{Fu2018jhep} that these phase transitions indicate that qubits may not exist on the RT surface and act entirely nonlocally. This nonlocality could be tracked in the dual theory using a bit thread picture, which may be employed to interpret more dual EWCS behavior \cite{Freedman2017cmp,Abt2018fp,Cui2020cmp}. In particular, the results show that the phase transition occurs at larger $s_c$ with larger $\alpha$. The profound link between this result and the bit threads in higher curvature gravity \cite{Harper:2018sdd} warrants further investigation.}

%%%%%%%%%%%%%%%%%%%%%%%%%%%%%%%%%%

\begin{figure}[h]
\centering
 \includegraphics[width=8cm]{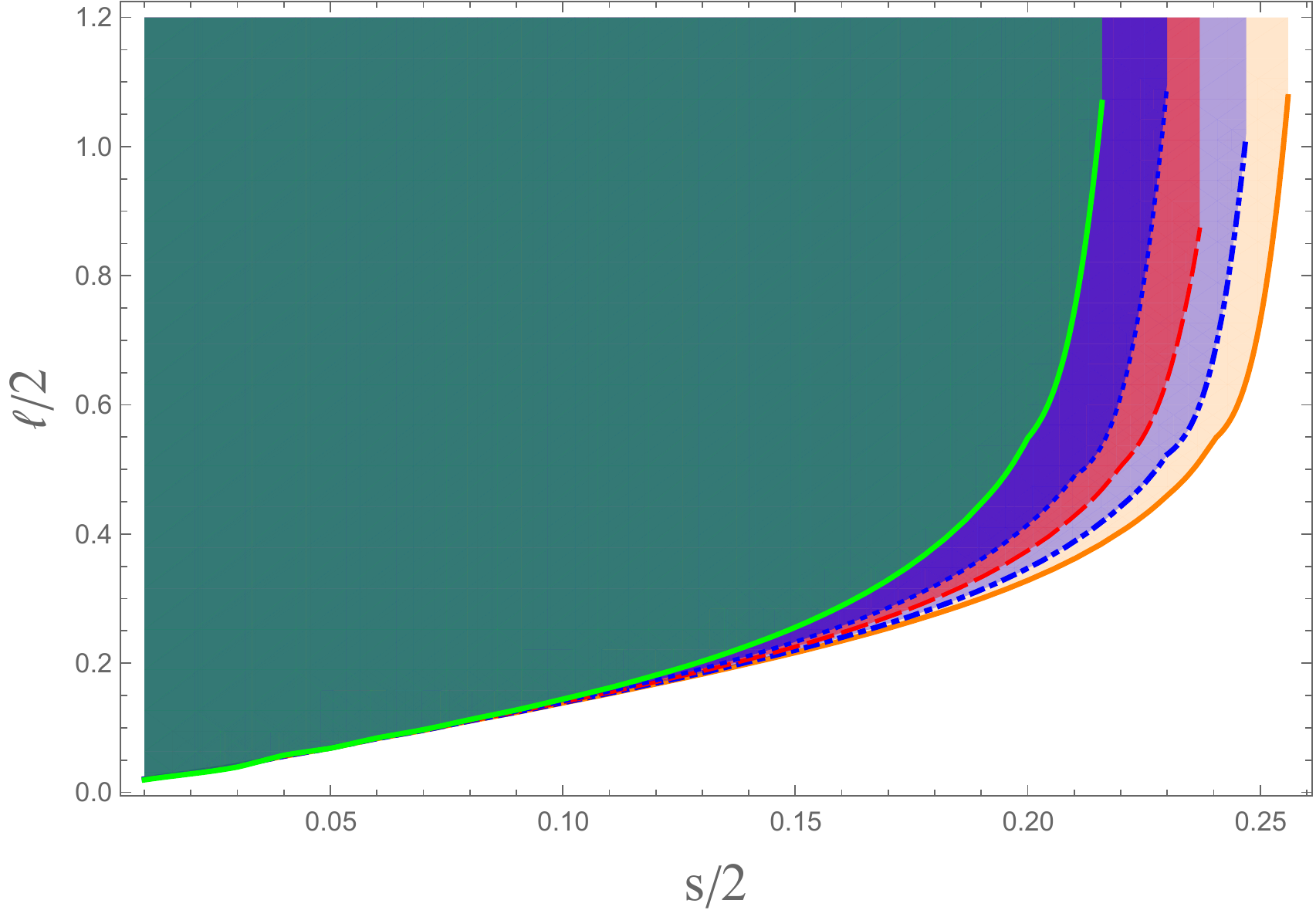}\ \hspace{0.1cm}
	  \caption{Nonzero region of RHMI for $\alpha=-0.19,\,-0.1 ,\,-0.05 ,\,0,\, 0.05\,$ from left to right. For fixed separation $s$, the RHMI shows a discontinuous phase transition when $\ell/2$ is smaller than the critical point on the curve. The end-point is roughly given on the right side of the critical curve due to the limitation of numerical calculation accuracy.}
 \label{fig:nzhmi}
\end{figure}

Further investigation is required to better demonstrate the effects of coupling constant on the discontinuous phase transition of RHMI (and thus REWCS). Figure~\ref{fig:nzhmi} shows the critical curves where RHMI is zero for different $\alpha$.
%%%
%%%%%
For fixed separation $s$ the nonzero RHMI and REWCS only exist for the shadow region above the critical curve based on the plot. When the critical curves with various $\alpha$ are compared, the curves are almost identical for small separation and width, but they become different as $s$ and $\ell$ increase. The parameters' range with nonvanishing REWCS is greater for larger $\alpha$.

Another feature shows that for each $\alpha$, there exists a unique $s_*$ where the two subregions disentangle when $s>s_*$, regardless of their width. Because of the limitations of our numerical method's accuracy, we can roughly confirm that \footnote{During the numerical calculation we have chosen the step size of separation as $\delta s=0.02$, which is not small enough to obtain an accurate numerical solution for $s_*$. But such a step size is good enough to describe the behaviors of RHMI and REWCS. These numerical results are consistent with Ref.\cite{Yang:2018gfq}.}
\begin{eqnarray}
\frac{s_{*}}{2}\simeq\left(0.256,\,0.247,\,0.237,\,0.231,\,0.216\right)\quad \text{for}\quad \alpha=\left(0.05,\,0,\,-0.05,\,-0.1,\,-0.19\right).
\end{eqnarray}

A relationship between $s_*$ and $\alpha$ can be observed. With the monotonicity of HEE, we assume that the strip width is so large that the area of the corresponding extremal surface can be represented as $\tilde{\mathcal{S}}(\Gamma_A)=\tilde{\mathcal{S}}(\tilde{\Gamma}_{A})+2\, \tilde{\mathcal{S}}(\Gamma_{0\rightarrow1})$ where $\tilde{\mathcal{S}}(\tilde{\Gamma}_{A})$ represents the area around the horizon $(z_h=1)$ and $\tilde{\mathcal{S}}(\Gamma_{0\rightarrow1})$ indicates radial subsection area given by Eqn.~(\ref{eopactionstatic}) with integral region $z\in[\epsilon,1]$. The area $\tilde{\mathcal{S}}(\Gamma_{ACB})$ can be expressed as $\tilde{\mathcal{S}}(\Gamma_{ACB})=2\left(\tilde{\mathcal{S}}(\tilde{\Gamma}_{A})+\tilde{\mathcal{S}}(\Gamma_{0\rightarrow1})\right)+\tilde{\mathcal{S}}(\tilde{\Gamma}_{C})$ where we use $\tilde{\mathcal{S}}(\tilde{\Gamma}_{C})$ to represent the area around the horizon for separation $C$ with width $s_*$. So the condition HMI$=0$ results in the following relation:
\begin{eqnarray}
\tilde{\mathcal{S}}(\tilde{\Gamma}_{C})+\tilde{\mathcal{S}}(\Gamma_{C})=2\, \tilde{\mathcal{S}}(\Gamma_{0 \rightarrow 1}).
\label{criticalseparation}
\end{eqnarray}
An expression for $\tilde{\Gamma}_{C}$ can be easily obtained. Since it indicates the area around the horizon corresponding to boundary width $s_*$, thus the boundary parameter is given by $x\in [-s_{*}/2, s_{*}/2]$, $z(x)=1$, $z'(x)=0$,
\begin{eqnarray}
\tilde{\mathcal{S}}(\tilde{\Gamma}_{C})=\int_{-s_{*}/2}^{s_{*}/2}dx\frac{1}{L_{AdS}^{3}}=\frac{s_{*}}{L_{AdS}^{3}}.
\end{eqnarray}

Unfortunately, $\tilde{\mathcal{S}}(\Gamma_{C})$ remain still a complicated function about $\alpha$ and $s_*$, whose analytical expression is difficult to obtain.

%%%%%%%%%%%%%%%%%%%%%%%%%%%%%%%%%%%%%%%%%%%%%%%%%%%%

\section{Entanglement wedge cross-section in Vaidya Gauss-Bonnet black brane}\label{sec:Vaidya result}

The evolution of EWCS is investigated in this section. In principle, the AdS/CFT theory theoretically requires a zero thickness limit of $v_0$ \cite{Balasubramanian:2010ce,Balasubramanian:2011ur}. Thus, the mass function, (\ref{massfunction}), should be chosen as a unit step function, resulting in $\partial_{v}f=0$ except at the point $v=0$. Assuming the Euler-Lagrange equation from the action (\ref{eopactiondy}) has a solution given by
\begin{equation}
Q(v,z(v))=\frac{dz}{dv}.
\label{simplesolution}
\end{equation}
Subtracting this solution into the Euler-Lagrange equation leads to
\begin{eqnarray}
&\left(f+2Q\right)\left(P_{1}-P_{2}\right)=0,\\\nonumber
&P_{1}=(f+Q)[2(d-1)\,f\,(f+2Q+2Q^2\,\alpha)-(f+2Q+6Q^{2}\,\alpha)\, z\, \partial_{z}f],\\ \nonumber
&P_{2}=[f+2Q+2\alpha\, Q\,(2f+3Q)]\,z\, \partial_{v}f.
\label{tempsolution1}
\end{eqnarray}
Theoretically {we have} $\partial_{v}f=0$ except for point $v=0$, thus the above equation holds with {the following condition}
\begin{equation}
f+2Q=0, \qquad \text{or} \qquad P_{1}=0.
\end{equation}
Notably, there is no ``and'' between two conditions since if they both hold, $f$ will no longer be a function of $v$.

Substituting the first condition into the action, one gets a zero denominator, indicating this condition does not hold. The second condition is actually the union of two subconditions,
\begin{equation}
f+Q=0, \quad \text{or} \quad 2(d-1)\,f\,(f+2Q+2Q^2\,\alpha)-(f+2Q+6Q^{2}\,\alpha)\, z\, \partial_{z}f=0.
\end{equation}
However, it is easy to check that the latter condition could never be fulfilled in our setup. Finally, the Euler-Lagrange equation has a solution given by the following:
\begin{equation}
-f(v,z(v))=\frac{dz}{dv}.
\label{simplesolution}
\end{equation}

With the mass function (\ref{massfunction}), the action (\ref{eopactiondy}) only gives a $z(v)$ solution that is close to the real extramal surface\footnote{Here, we use the word ``real'' to represent the extremal surface given by the limit $v_0\rightarrow 0$.}, as does Eqn.~(\ref{simplesolution}). As $v_0\to 0$, the contribution of the points near $v=0$ to the HEE \eqref{eopactiondy} could be negligible because the integral function around $v=0$ is finite. Smaller $v_0$ results in a $z(v)$ that is closer to the real extremal surface. Hereafter, we argue that using solution (\ref{simplesolution}) as an approximate extremal surface is beneficial. Finally, the holographic EoP can be expressed as follows:
\begin{equation}
\tilde{S}^{dy}_{eop}(V)=\int_{z_{*}^{b}}^{z_{*}^{u}}\frac{dz}{z^{d-1}L_{AdS}^{d-2}}\left[\sqrt{f(v,z)^{-1}}+2 \alpha \sqrt{f(v,z)}\right].
\label{finalaction}
\end{equation}
Again, both $z_{*}^{b}$ and $z_{*}^{u}$ should be calculated at the same boundary time.

\begin{figure}[h]
\centering
 \includegraphics[width=\textwidth]{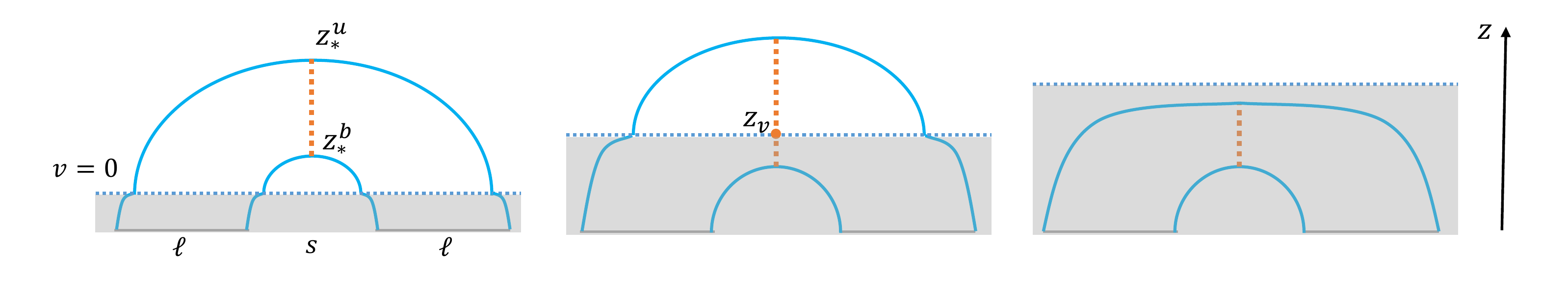}%\ \hspace{0.1cm}
	  \caption{Three stages to calculate entanglement wedge cross-section according to the shell position $z_{v}$. (Left) the first stage with $0<z_{v}<z_{*}^{b}$;  (Middle) the second stage with $z_{*}^{b}<z_{v}<z_{*}^{u}$; (Right) the third stage with $z_{*}^{u}<z_{v}$; For all three panels, the dotted horizontal line indicates the shell position. The shadowed region stands for AdS-GB black hole spacetime, i.e., $v>0$. The unshadowed region represents the pure AdS-Gauss-Bonnet spacetime with $v<0$. The dotted vertical orange line indicates the extremal surface for calculating EWCS.}
 \label{fig:threestage}
\end{figure}

If the radial position of the shell is labeled as $z_{v}=z_{v}(t)$, then the radial EoP, in this case, should be expressed according to $z_v$, see Figure \ref{fig:threestage}.
%%%%%%
\begin{itemize}
\item[1).] For $0<z_v<z_{*}^{b}$, the shell has not {reached} the point $z_{*}^{b}$; thus the EoP should be given by the following:
\begin{eqnarray}
\text{EoP}_{1}=\int_{z_{*}^{b}}^{z_{*}^{u}}\frac{dz}{z^{d-1}L_{AdS}^{d-3}}\left(1+\frac{2\alpha}{L_{AdS}^{2}}\right);
\label{EWt1}
\end{eqnarray}
\item[2).] For $z_{*}^{b}<z_v<z_{*}^{u}$, the location of shell is between ($z_{*}^{b},\, z_{*}^{u}$), and the EoP is given by the following:
\begin{eqnarray}
\text{EoP}_{2}=\int_{z_{*}^{b}}^{z_v}\frac{dz}{z^{d-1}L_{AdS}^{d-2}}\left[\sqrt{f(z)^{-1}}+2 \alpha \sqrt{f(z)}\right]+\int_{z_v}^{z_{*}^{u}}\frac{dz}{z^{d-1}L_{AdS}^{d-3}}\left(1+\frac{2\alpha}{L_{AdS}^{2}}\right);
\label{EWt2}
\end{eqnarray}
\item[3).] For $z_{*}^{u}< z_v$, the system has reached the equilibrium, and the EoP is directly given by Eqn.~(\ref{eopactionstatic}).
\end{itemize}

We first represent the qualitative evolutionary of the unrenormalized HMI and EWCS. By numerical method, we set the initial conditions as follows:
\begin{eqnarray}
z(0)=z_{*},\quad v(0)=v_{*},\quad z(\pm\frac{\ell}{2})=z_{0},\quad v(\pm\frac{\ell}{2})=t_0,\quad
z'(0)=v'(0)=0,
\label{initialconditiondy}
\end{eqnarray}
where $z_0$ is the UV cut-off and $t_0$ is the boundary time. Hereafter, we set $z_0=0.02$. The equations of motion are given by Eqn.~(\ref{EEaction2}) with $f(z)$ replaced with $f(z,v)$. The coupling constant is chosen as $\alpha=\pm0.05,\,\pm0.02,\,\pm 0.01,\,0,$.
%%%%%%%%%%%%
\begin{figure}
\centering
 \includegraphics[width=\textwidth]{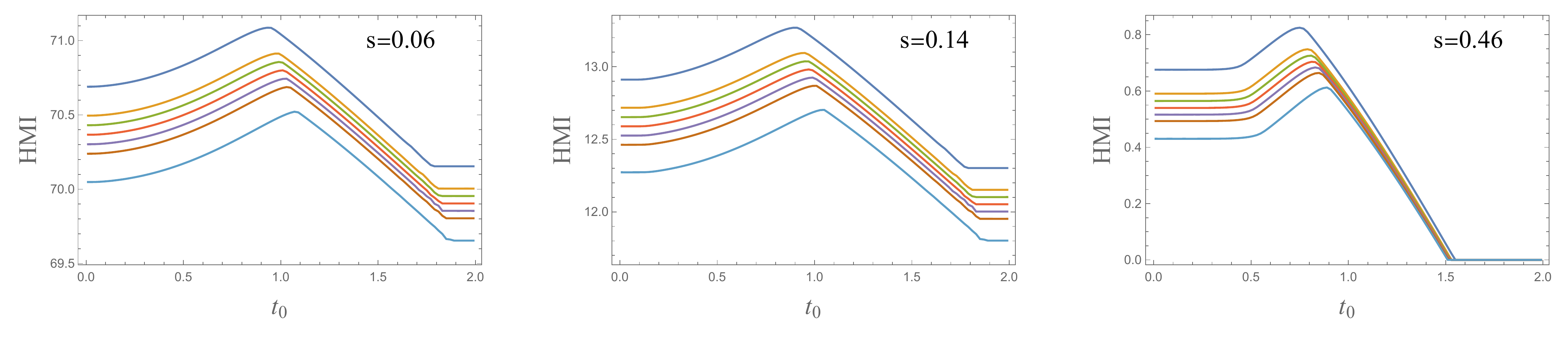}%\ \hspace{0.1cm}
	  \caption{The evolutions of holographic mutual information for different $\alpha$ with fixed $s=0.06,\,s=0.14,\,s=0.46$
 (from left to right), respectively. For each panel, from top to bottom the coupled constant $\alpha$ is chosen as $\alpha=0.05,\, 0.02,\,0.01,\,0,\,-0.01,\,-0.02,\,-0.05$. Further, we have rescaled HMI in the first and second panel of this figure as $\text{HMI}=\text{HMI}_{\alpha}-(\text{HMI}^{cons}_{\alpha}-\text{HMI}^{cons}_{-0.05})+5\alpha$ to make the curves more concise and comparative without losing details. $\text{HMI}_{\alpha}$ indicates the nonrescaled values of HMI with the coupled constant $\alpha$, whereas $\text{HMI}^{cons}_{\alpha}$ is its equilibrium value. }
 \label{fig:HMIt0}
\end{figure}
%%%%%
\begin{figure}
\centering
 \includegraphics[width=\textwidth]{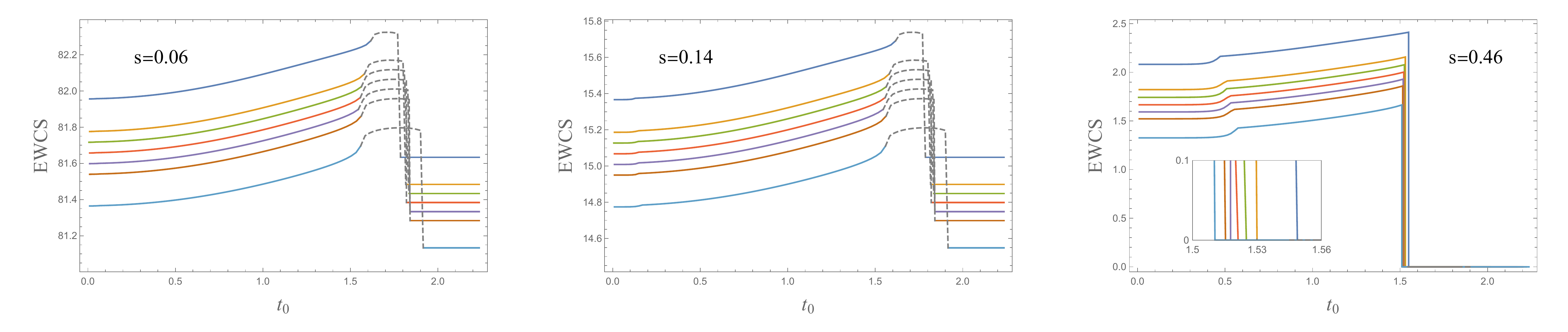}%\ \hspace{0.1cm}
	  \caption{The evolution of entanglement wedge cross-section for different $\alpha$ with fixed $s=0.06,\,s=0.14,\,s=0.46$ (from left to right). For each panel, from top to bottom the coupling constant $\alpha$ is chosen as $\alpha=0.05,\, 0.02,\,0.01,\,0,\,-0.01,\,-0.02,\,-0.05$. We have also rescaled EWCS in the first and second panels as $\text{EWCS}=\text{EWCS}_{\alpha}-(\text{EWCS}^{cons}_{\alpha}-\text{EWCS}^{cons}_{-0.05})+5\alpha$ (Figure \ref{fig:HMIt0}) to make the curves more comparative without losing details. $\text{EWCS}_{\alpha}$ indicates the nonrescaled values of EWCS with the coupled constant $\alpha$, whereas $\text{EWCS}^{cons}_{\alpha}$ is its equilibrium value.}
 \label{fig:EWCSwitht0}
\end{figure}
%%%%%%%%%%%
\begin{figure}
\centering
 \includegraphics[width=0.85\textwidth]{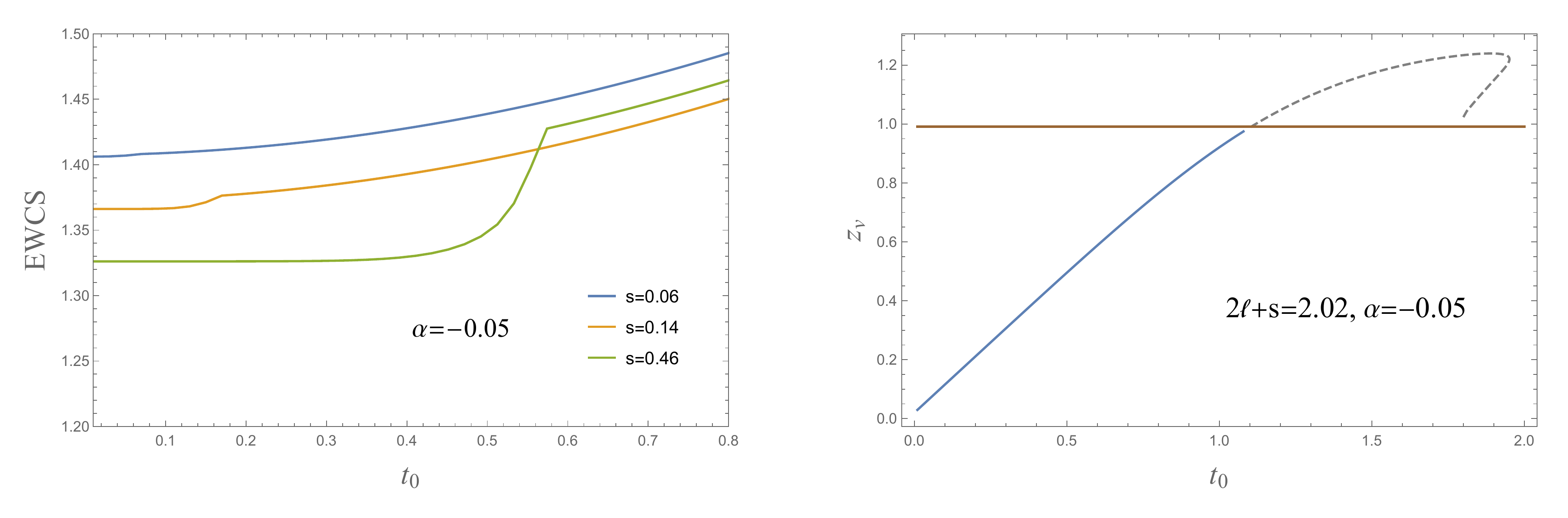}%\ \hspace{0.1cm}
	  \caption{(Left) The evolutions of entanglement wedge cross-section at early stage for $\alpha=-0.05$. To compare the different behaviors of three cases ($s=0.06, 0.14 and 0.46$), we have rescaled the EWCS by reducing a constant for each curve without changing its behavior. (Right) The evolution of the shell position for $\alpha=-0.05$ with $2\ell+s=2.02$ .
}
 \label{fig:zvevolution}
\end{figure}

Figure \ref{fig:HMIt0} presents the HMI evolution with various strip separations, $s$\footnote{As shown in the caption, we will choose three values for the strip separation, i.e., $s=0.06,\,s=0.14,\,s=0.46$. Here we shall interpret why we choose those values. The choice of the smallest value $s=0.06$ is because our numerical precision requires the minimal separation in the calculation to be $s/2\simeq0.03$; otherwise, the code would fail. $s=0.46$ was chosen because the phase transition occurs at $s/2\approx 0.23$, and for larger $s$, the two stipes would disentangle as we will presented shortly. Then the middle value $s=0.14$ between $s=0.06$ and $s=0.46$ was arbitrarily chosen; however, this did not affect our representation on the EWCS evolution.}, showing a monotonic behavior between HMI and the coupling constant $\alpha$. {Apparently, for sufficiently large separation $s$, the evolution of EWCS could be divided into three stages, see Figure \ref{fig:threestage}. In the first stage, the shell has not reached $z_{*}^{b}$ yet. The EWCS starts with its vacuum value and continues until the shell reaches $z_{*}^{b}$ \cite{Li:2013sia,Tanhayi2016jhep}. The HMI increases linearly to an upper limit as the entanglement entropy associated with strips A (and B) saturates to its equilibrium value after the shell advances into the second stage. Afterward, HMI decreases linearly until all the entanglement entropy saturates to their equilibrium value, i.e., the shell passes the tipping point $z_{*}^{u}$. These features of HMI were also observed in Einstein gravity \cite{Allais2012jhep,Alishahiha2015jhep}. Especially, the first two panels show that the equilibrium time {increases as $\alpha$ decreases from $0.05$ to $-0.05$}, whereas the third panel exhibits the contrary behavior. We will demonstrate shortly how such behaviors are closely linked to EWCS involvement and the interesting dependence of EWCS on the GB coupling.}
%MMMMMMM
%In all cases, HMI grows as the time involves at first, then decreases to a constant indicating the system reaches equilibrium, and the constant is always smaller than the initial value. With time evolving, for small separations the HMI is always positive, meaning that two strips are always entangled till the equilibrium (see the left and middle panels); while for large enough separations, the HMI finally drops from positive values to be zero, meaning that the two strips are initially entangled then being disentangled (see the right panel).  and gravity with Hyperscaling violation \cite{Tanhayi2016jhep}. We will not expand the details about such dependent behaviors of the involving of HMI on $\alpha$ because they have been carefully discussed in pioneering literature \cite{Li:2013sia}. }

The evolution of EWCS for different $\alpha$ with fixed separations is shown in Figure \ref{fig:EWCSwitht0}.
{ It is obvious that for the left and middle panels, the strips are entangled with small separations during the thermal quench, whereas for the right panel, the two strips are only entangled before a transition time for an adequate range of $s$. This observed feature from the EWCS behavior is expected based on Figure \ref{fig:HMIt0}. Further, more universal features could also be extracted from the EWCS evolution in the plots: In the first stage, EWCS behaves as a constant and the left panel of Figure \ref{fig:zvevolution} shows this early-stage constant behavior by using $\alpha=-0.05$ as an example. This is because the spacetime in this stage is pure AdS with GB corrections as shown in Figure \ref{fig:threestage}. As time evolves to the second stage, EWCS monotonously increases and migrates to a ``numerical deformation region'' (gray dashed lines), which will be explained later. Finally, EWCS becomes a constant again as the system reaches thermal saturation at a transition time, and this constant could be nonzero or zero depending on the separation. These behaviors are consistent with the EWCS evolution in Einstein gravity \cite{BabaeiVelni:2020wfl} and strongly relates to the HMI evolutionary features. Moreover, such a transition time increases as $\alpha$ {decreases} in each plot, implying that the {smaller} GB parameter suppresses the thermalization, which {is consistent with the evolution of HEE in Refs. \cite{Shao2015PRD,Farsam2019GRG}.}% but different with \cite{Li:2013cja,Caceres2017jhep}.}

To explain the {numerical deformation region} during the evolution, we plotted the shell position as time evolves by using $\alpha=-0.05$ as an example in the right panel of Figure \ref{fig:zvevolution}. The boundary region is chosen as $2\ell+s=2.02$. The brown line indicates the equilibrium $z^{equ}_{*}$ of the extreme surface, i.e., the turning point of $\Gamma_{ACB}$ in Figure \ref{fig:eop}. The curve of $z_v$ intersects $z_{*}^{equ}$ at $t_0\approx1.2$, which further enters the {deformation} region. Mathematically, the shell falls into ``the horizon'' after $z_v\geq1$ but the thermal equilibrium is not yet to be attained. We argue that this behavior results from the adopted thermalization model, i.e., the rectangular strips and the shell with small thickness. And it will disappear if we calculate the holographic entanglement entropy using the shock wave (zero-thickness limit) or other boundary shapes. {Appendix B} presents a detailed numerical and theoretical explanation.

One may note that the time when $z_v$ intersects $z_{*}^{equ}$ is different from when the EWCS curve enters the deformation region. This is due to the selected mass function $m(v)$. Theoretically, the shell should be zero thickness with certain mass, as the energy injects instantaneously. But we have to choose $v_0\neq0$ in conventional numerical calculations. In our case, $v_0=0.02$, thus using Eqn. (\ref{massfunction}) as mass function gives us $m(v\leq-0.05 )\approx0$ and $m(v\geq0.05)\approx 1$. However, $v=0$ indicates the position of the shell, so $m(0)=1/2$, making $f(z,v)$ a real valued function for $z\in[z_0,1.1892]$. Thus, when $z_v\simeq 1.1892$, $t_0\simeq 1.58$ is given, causing two time points to coincide.

We {have also investigated} the relationship between EWCS and the separation $s$ at different time slices $t_0=0.01, 1.01, 1.51 and 1.99$ for different $\alpha$, see Figure \ref{fig:EWCSwiths}. {In each plot, as the separation becomes wider, {EWCS} becomes smaller and suddenly drops to be zero at certain $s_{*}$. This behavior is similar to that in the static case shown in the right plot of Figure \ref{fig:rhmi}, below which we have argued this behavior as a phase transition. Thus, we shall analyze the effect of $\alpha$ at different times in detail.}
%%%%%%
\begin{figure}[h]
\centering
 \includegraphics[width=0.85\textwidth]{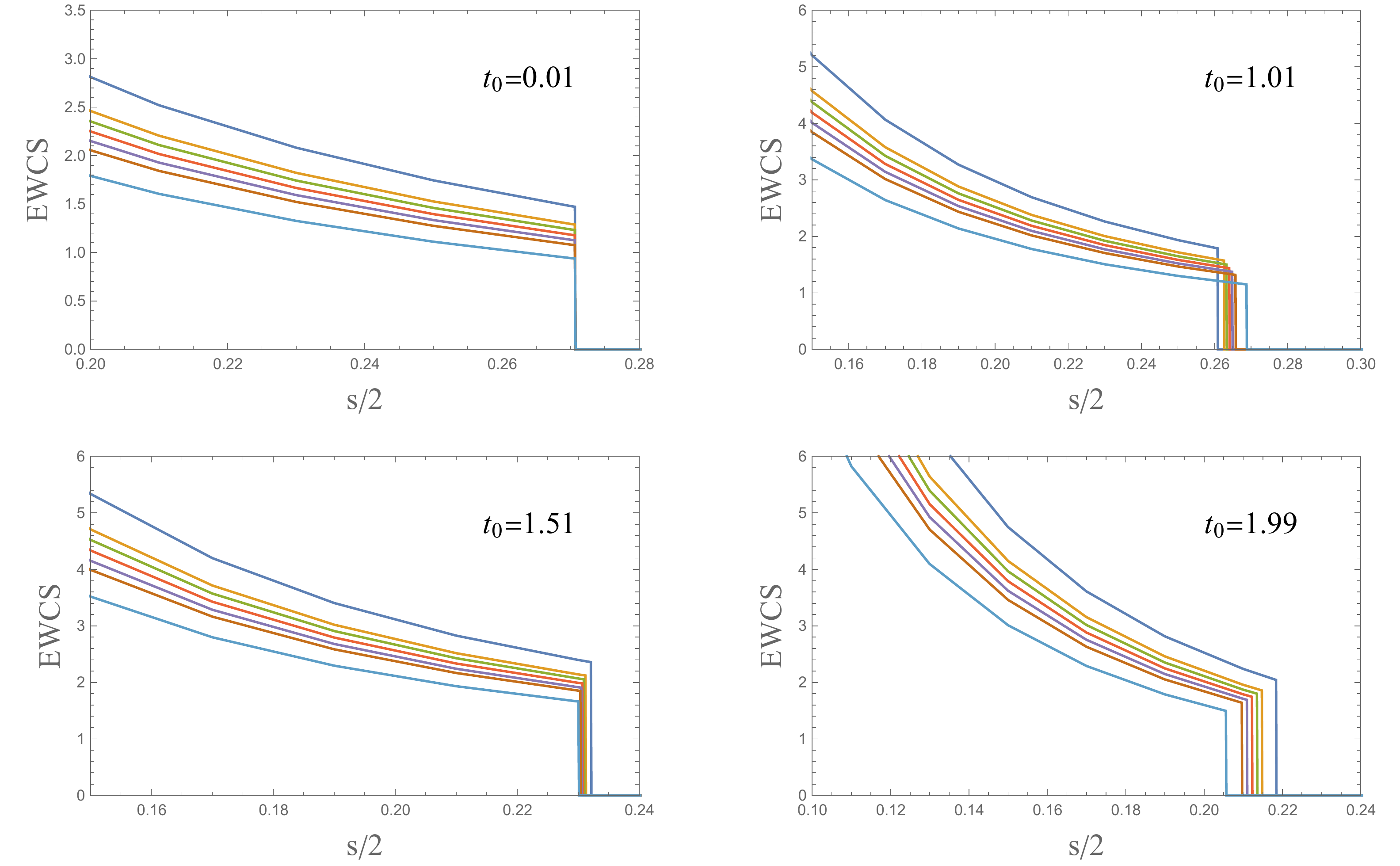}%\ \hspace{0.1cm}
	  \caption{The entanglement wedge cross-section with the strip width for different $\alpha$ at different times. For each panel, from top to bottom  $\alpha=0.05,\, 0.02,\,0.01,\,0,\,-0.01,\,-0.02,\,-0.05$.}
 \label{fig:EWCSwiths}
\end{figure}

In the first case, $t_0=0.01$ indicates the beginning of the thermal quench, when the spacetime is pure AdS with GB corrections. The metric (\ref{BH}) can then be rescaled as
\begin{equation}
ds^{2}=-\tilde{r}^{2}dt^{2}+L_{AdS}^{2}\frac{d\tilde{r}^{2}}{\tilde{r}^{2}}+\tilde{r}^{2}d{\mathbf{x}}^{2},
\label{BH2}
\end{equation}
where $\tilde{r}^{2}=r^{2}/L_{AdS}^{2}$. Such rescaling suggests that the GB corrections will not affect the phase transition of HMI in the pure AdS-GB spacetime since $L_{AdS}^2$ will be eliminated during the calculation. Consequently, for different $\alpha$, EWCS will transition from the entangled phase to disentangled phase at the same separation $s$.

For cases {with} $t_0=1.01$ and $t_0=1.51$, the EWCS is divided into two parts as the shell falls. The significant difference is that the transition separations $s_*$ are no longer monotonic with $\alpha$, which is consistent with the evolutionary properties of HMI depicted in Figure \ref{fig:HMIt0}. For example, when $s=0.46$, the HMI becomes zero for different $\alpha$ around $t_{0}\approx 1.5$. Whereas, when $t_{0}=1.51$, the EWCS becomes zero around $s/2\approx0.23$.

This behavior can be explained as follows. In the currently adopted physical framework, we claim that the decreasing rate of the transition separations $s_*$ is consistent with $\alpha$ until a specified boundary time $t^{*}_{0}$ is reached. Then the decreasing rate is contrary to the value of $\alpha$ until the entire system enters the thermal equilibrium with fixed $s_*$. Such a boundary time should be larger enough so that for the two subregions $A$, $B$ and separation $C$ are all in the thermal equilibrium state, but not for the combined region $ACB$, as shown in the middle panel of Figure \ref{fig:threestage}. Since the transition separations $s_{*}$ should be given by the relation \eqref{RHMI} with $I(A,B)=0$, then one will have

\begin{equation}
2\mathcal{S}^{st}(\ell)=\mathcal{S}^{dy}(2\ell+s_{*})+\mathcal{S}^{st}(s_{*}),
\label{eqtransition}
\end{equation}
where $''st''$ and $''dy''$ represent the static and dynamic cases, respectively. In our dynamic setup $2\ell+s=2.02$ is fixed, so one must have $s<\ell<0.675$. $\mathcal{S}^{dy}(2\ell+s)$ is not only larger for larger $\alpha$, but is also a monotonous function of the boundary time, which explains the decrease in $s_{*}$ with the boundary time. Suppose $\alpha_{2}>\alpha_{1}$, then at a boundary time $t_{1}<t_{0}^{*}$, one has the following:
\begin{eqnarray}
2\mathcal{S}_{\alpha_{1}}^{st}(\ell_{\alpha_{1}})=\mathcal{S}_{\alpha_{1}}^{dy}(2\ell_{\alpha_{1}}+s_{*\alpha_{1}})+\mathcal{S}_{\alpha_{1}}^{st}(s_{*\alpha_{1}}), \label{specifieds1}\\
2\mathcal{S}_{\alpha_{2}}^{st}(\ell_{\alpha_{2}})=\mathcal{S}_{\alpha_{2}}^{dy}(2\ell_{\alpha_{2}}+s_{*\alpha_{2}})+\mathcal{S}_{\alpha_{2}}^{st}(s_{*\alpha_{2}}).
\label{specifieds2}
\end{eqnarray}
{Taking} (\ref{specifieds1})-(\ref{specifieds2}), one derives
\begin{equation}
2 \Delta\mathcal{S}^{st}_{\ell}=\Delta\mathcal{S}^{dy}+\Delta\mathcal{S}^{st}_{s_*},
\label{specifieds3}
\end{equation}
where
\begin{eqnarray}\nonumber
\Delta\mathcal{S}^{dy}&=&\mathcal{S}_{\alpha_{2}}^{dy}(2\ell_{\alpha_{2}}+s_{*\alpha_{2}})-\mathcal{S}_{\alpha_{1}}^{dy}(2\ell_{\alpha_{1}}+s_{*\alpha_{1}}),\\ \nonumber
\Delta\mathcal{S}^{st}_{\ell}&=&\mathcal{S}_{\alpha_{2}}^{st}(\ell_{\alpha_{2}})-\mathcal{S}_{\alpha_{1}}^{st}(\ell_{\alpha_{1}}),\\ \nonumber
\Delta\mathcal{S}^{st}_{s_*}&=&\mathcal{S}_{\alpha_{2}}^{st}(s_{*\alpha_{2}})-\mathcal{S}_{\alpha_{1}}^{st}(s_{*\alpha_{1}}). \nonumber
\end{eqnarray}
Generally speaking, {if $\alpha_{1}$ and $\alpha_{2}$ correspond to the same transition separation, } then one should have $\Delta\mathcal{S}^{st}_{s_*}=\Delta\mathcal{S}^{st}_{\ell}=\Delta\mathcal{S}^{dy}$. However, in our case, the numerical calculations show that for same boundary strip size, $\mathcal{S}_{\alpha_{2}}^{st}(s_*)-\mathcal{S}_{\alpha_{1}}^{st}(s_*)<\mathcal{S}_{\alpha_{2}}^{st}(\ell)-\mathcal{S}_{\alpha_{1}}^{st}(\ell)<\Delta\mathcal{S}^{dy}$ (as $s_{*}$ must be smaller than $\ell$). Combined with the facts that HEE increases monotonously with the strip size or the boundary time, so Eqn. (\ref{specifieds3}) holds only when $\Delta\mathcal{S}^{st}_{\ell}$ is large enough but $\Delta\mathcal{S}^{st}_{s_*}$ is small enough\footnote{Note in this case $\Delta\mathcal{S}^{st}_{\ell}$ must be larger than $\Delta\mathcal{S}^{dy}$. Otherwise Eqn. (\ref{specifieds3}) will no longer hold.}. This means that a smaller $s_*$ is required for larger $\alpha$. As the boundary time evolves to $t_2>t_1$, $\Delta\mathcal{S}^{dy}$ becomes slightly larger, causing an additional difference $\delta\mathcal{S}^{dy}(t_2)=\Delta\mathcal{S}^{dy}(t_2)-\Delta\mathcal{S}^{dy}(t_1)>0$ on the right hand side of Eqn. (\ref{specifieds3}) compared with the case of time $t_1$. This very tiny difference could be used to balance both sides of Eqn. (\ref{specifieds3}), making $\Delta\mathcal{S}^{st}_{s_*}$ slightly larger and $\Delta\mathcal{S}^{st}_{\ell}$ slightly smaller. So in this case, one need a small $s_{*\alpha_1}$ and a large $s_{*\alpha_{2}}$. Especially, there must exist a special boundary time $t_{0}^{*}$ so that
\begin{equation}
\Delta\mathcal{S}^{st}_{\ell}(t_{0}^{*})-\delta\mathcal{S}^{dy}(t_{0}^{*})/3=\Delta\mathcal{S}^{dy}(t_1)=\Delta\mathcal{S}^{st}_{s_*}(t_{0}^{*})+\delta\mathcal{S}^{dy}(t_{0}^{*})/3,
\end{equation}
where $\Delta\mathcal{S}^{st}_{\ell}(t_{0}^{*})=\mathcal{S}_{\alpha_{2}}^{st}(\ell, t_{0}^{*})-\mathcal{S}_{\alpha_{1}}^{st}(\ell, t_{0}^{*})$ and $\Delta\mathcal{S}^{st}_{s_*}(t_{0}^{*})=\mathcal{S}_{\alpha_{2}}^{st}(s_*, t_{0}^{*})-\mathcal{S}_{\alpha_{1}}^{st}(s_*, t_{0}^{*})$.

As time {involves} the system reaches equilibrium, becoming the static case discussed in Section \ref{sec:static result}. Comparing the fourth panel ($t_0=1.99$) with the right panel of Figure \ref{fig:rhmi}, the relationship among EWCS, $s$ and $\alpha$ are same, with only numerical discrepancies. Such discrepancies result from the different numerical initial conditions we selected in the static and dynamic cases, which does affect the final discussions.

\section{Conclusions and Discussions}

To explore the properties of entanglement of purification with higher-order corrections, we studied its holographic dual, i.e., the EWCS using the AdS/CFT correspondence and the GB-Vaidya model. For simplification, two identical and separated rectangular strips in the boundary CFT are considered, and thus only the radial EWCS is concerned. Further, we investigated the characteristics of EWCS and its relationship with the GB coupling constant $\alpha$ in static and dynamics cases using the numerical method.

In the static case, the calculations suggest a monotonic relationship between the REWCS and $\alpha$, i.e., the larger $\alpha$, the larger REWCS. In addition, for large $\alpha$, the phase transition between the entangling and disentangling phases occur at large separation between two strips. Note the entanglement measure in the pure state is related to (H)RT surfaces, whereas it is described by EWCS in mixed state, which could be considered as the generalization of (H)RT surface. Thus, the different effects of GB correction on the (H)RT surfaces and EWCS from the bulk side warrant further investigation from an analytical perspective, since our findings are from numerical investigations. Further, understanding the deeper physics brought about by higher curvature correction from the boundary field theory side is fascinating, see possible directions on the discussion of higher curvature gravity from entanglement using the perturbative nature of conformal field theories \cite{Haehl:2017sot} and bit threads in higher curvature gravity \cite{Harper:2018sdd}.

For the dynamic case, we explored the evolution of EWCS using the holographic thermalization procedure. The monotonic relationship between EWCS and $\alpha$ holds during the entire evolution process. Nonetheless, the separations where the phase transition occurs are nonmonotonic with $\alpha$, and can be summarized in three stages. First, at the beginning of thermalization, the phase transition occurs at the same separation for all $\alpha$. In the second stage, the entangled strips disentangle at small separations for large $\alpha$. Finally, in the third stage, the entangled strips transform into the disentangled phase at large separations for large $\alpha$ which is consistent with that in the static case. These behaviors indicate that the higher-order corrections have nontrivial effects on the entanglement of purification. Especially, we have observed a nonmonotonic relationship between the transition separations and $\alpha$ for the evolution of EWCS. This behavior indicates that during the thermal quench, two discrete subregions on the boundary strongly coupled field disentangle at small separation with large $\alpha$ in the early time, and then disentangle at large separation with large $\alpha$ in the later period.

Inspired by the pioneering work \cite{Balasubramanian:2010ce,Balasubramanian:2011ur}, two-point functions, Wilson loops and entanglement entropy have been used in terms of their dual geometric objects to probe the thermalization process, which is related to the process of black hole formation in higher order derivative gravity (for example, see \cite{Caceres2017jhep} and reference therein). Inspired by this proposal, the evolution of more correlation quantities in CFT have been holographically investigated during the thermal quench process in general gravity theory, for instance, subregion complexity via complexity=volume conjecture \cite{BChen2018jhep,Ling:2018xpc,Cai2018prdc,yuting2019prd,Ling:2019ien}, entanglement of purification dual to EWCS  and complexity of  purification dual to the volume between the boundary and the EWCS surface, as stated in the introduction. Several remarkable evolutionary properties of these correlation quantities have been explored in various bulk theories, which provide the opportunity for us to further understand holography. Here for the sake of simplicity, we shall briefly compare the features of HEE, HMI, EWCS and complexity in GB gravity. By definition, the first three quantities are closely interrelated.

\begin{itemize}
\item{The phase transition from the connected to the disconnected phase occurs simultaneously for EWCS and HMI with the same GB coupling constants and strip separation. From Figure \ref{fig:threestage}, if the two strips are kept in a connected phase, EWCS and HMI will saturate simultaneously for {the} same $\alpha$ but the HEE of two strips will saturate earlier. }
\item{In our case, both EWCS and HMI are monotonic to $\alpha$, i.e., for a larger $\alpha$, one will have larger EWCS and HMI if they are in the connected phase with fixed separation and boundary time. However, some literature has shown that larger $\alpha$ may lead to smaller HEE \cite{Li:2013cja,Tanhayi2018epjc}. Although, we can check that HMI still satisfies the aforementioned monotonicity.}
\item{Different conjectures can be used to calculate the complexity in holography. For the  complexity=volume (CV) and CV2.0 conjecture, it has been found that the GB term suppresses the growth rate  \cite{Cai2018prdc}. In other words, larger $\alpha$ results in a longer saturation time. This is contrary to EWCS and HMI scenarios. Exploring the physical origination of the difference will be fascinating.}
\end{itemize}

Thus far, the dual two-point functions, Wilson loops, entanglement entropy and MI with GB corrections during the thermal quench have been numerically studied in \cite{Li:2013cja,Li:2013sia,Zeng:2013mca,Zeng2014jhep}. It was found that in GB theory, the relationship between the critical thermalization time and boundary region encounters certain ``phase transition", which could introduce the nontrival EWCS evolution relative to $\alpha$ due to the phase transition as shown in Figure \ref{fig:EWCSwiths}. The physical origins of this phase transition may be attributed to van der Waals-like phase transitions of the thermal entropy, which implies that below a critical GB parameter, the physically favorable thermal phase could be composed of a small or large black hole
\cite{Song2016npb}. Particularly, the physically {favorable} GB black hole could jump from the small to the large one during the evolution depending also on $\alpha$, resulting in nontrivial behaviors of EWCS with respect to GB coupling at different times. It was also found in \cite{Song2016npb} that Wilson loop, holographic entanglement entropy and two-point correlation function all {exhibit} van der Waals-like phase transitions as the thermal entropy. Affirmatively, in-depth physics on the nontrivial effect of $\alpha$ on the correlation quantities and their evolution requires further investigation.
Moreover, we hope to generalize the entanglement tsunami picture \cite{Liu:2013qca} to analytically study the evolving properties of EWCS in different scaling regimes during the thermalization process, which could help understand those interesting properties.

Besides the main results of this paper, we also noted that a {deformation} region appears in the evolution of EWCS when the separation is small but $2\ell+s$ is large enough. We have argued that such behaviors are caused by the selected model. Checking the influence of different boundary regions on the evolution of EWCS or HMI with GB corrections or improving the physical model may provide insight into the deep physical prescription of EoP evolution. It is noticed that we focused on the symmetric case with two identical subsystems in this study. It would be fascinating to see if the rule still holds in an asymmetric condition. We intend to address this problem in the future after improving numerical skills.

\begin{acknowledgments}
This work is partly supported by the Natural Science Foundation of China under Grants Nos. 11947067, 12005077 and 11705161. Y-Z. Li is also supported by fund No.1052931902 for doctoral research of Jiangsu university of science and technology. X-M. Kuang is also supported by Fok Ying Tung Education Foundation under Grant No.171006 and  Natural Science Foundation of Jiangsu Province under Grant No.BK20211601.
\end{acknowledgments}

\section*{Appendix A: correlation measures connected with the EWCS}\label{appendixA}
In this appendix, we shall briefly review the definitions of entanglement of purification ($E_P$), reflected entropy($S_R$), odd entropy($S_O$) and logarithmic negativity($\xi$). Let us consider two subsystems $A$ and $B$ in the Hilbert space $\mathcal{H}_A$ and $\mathcal{H}_B$, respectively. In addition, $\rho_{AB}$ is the mixed density matrix for
$A\cup B$ living in the total Hilbert space $\mathcal{H}=\mathcal{H}_A\otimes\mathcal{H}_B$.

\begin{itemize}
  \item Entanglement of purification

$E_P$ is a measure of the total (classical and quantum) correlations between two
subsystems. To compute this quantity, some auxiliary degrees of
freedom will be added to $\mathcal{H}$, after which  the total enlarged Hilbert space becomes $\bar{\mathcal{H}}=\mathcal{H}_A\otimes\mathcal{H}_B\otimes\mathcal{H}_A\otimes\mathcal{H}_B$. Further, the mixed state can be purified by constructing a pure state $\mid \psi\rangle\langle\psi\mid$  such that $\rho_{AB} = \mathrm{Tr}_{A'B'}\mid \psi\rangle\langle\psi\mid$ and
$\mid\psi\rangle \in \mathcal{H}_{AA'}\otimes \mathcal{H}_{BB'}$, though this purification is not unique. Then $E_P$ is defined by the minimum EE between $A$ and its auxiliary partner $A'$ for all possible purifications, which are \cite{Terhal:eop}
\begin{equation}
E_P(A,B)=\mathrm{MIN}_{\rho_{AB}=\mathrm{Tr}_{A'B'}(\mid \psi\rangle\langle\psi\mid)}S(A,A').
\end{equation}
We observed that the above definition recovers EE when $\rho_{AB}$ is pure.

  \item Reflected entropy

$S_R$ is a new measure of the total correlation between two disjoined subsystems. To define $S_R$, we could double the initial Hilbert space $\mathcal{H}$ to $\mathcal{H}\otimes\mathcal{H}'$, and then canonically purify the mixed state, $\rho_{AB}=\Sigma_i P_i\mid\rho\rangle\langle\rho\mid$, such that $\sqrt{\rho}=\Sigma_i\sqrt{P_i}\rho_i\otimes\rho_i$ is a pure state in  $\mathcal{H}\otimes\mathcal{H}'$. Then $S_R$
is defined as the EE between $A$ and $A'$ as \cite{Dutta:2019gen}
\begin{equation}
S_R(A,B)=-\mathrm{Tr}(\rho_{AA'}\ln\rho_{AA'}),
~~~~~\sqrt{\rho_{AA'}}=\mathrm{Tr}_{BB'}\mid\sqrt{\rho}\rangle\langle\sqrt{\rho}\mid.
\end{equation}
It is obvious that the above reflected entropy also reduces to EE if $\rho_{AB}$ is {the one for} pure state.

  \item Odd entropy

$S_O$ is also a new measure of correlations for the mixed states. It is defined as follows \cite{Tamaoka:2018ned}:
\begin{equation}
S_O(A,B)=\mathrm{Lim}_{n_O\to1}\frac{1}{1-n_O}\left(\mathrm{Tr}(\rho_{AB}^{TB})^{n_O}-1\right)
\end{equation}
where $\rho_{AB}^{TB}$ denotes the partial transpose of  $\rho_{AB}$ with respect to $B$.
It was proved that similar to the $E_P$ and $S_R$, $S_O$ also reduces to the EE  when the state is pure.

  \item Logarithmic negativity

The logarithmic negativity captures only the quantum correlations for mixed states unlike the purification entanglement, reflected entropy, and odd entropy. It is definition is as follows \cite{Plenio:2005cwa}:
\begin{equation}
\xi(A,B)=\ln\mathrm{Tr}(\rho_{AB}^{TB}).
\end{equation}

\end{itemize}

\section*{Appendix B: more explanations for the numerical deformation region}\label{appendixB}
In this appendix, we will provide more numerical and theoretical explanations for the numerical deformation region in Figure \ref{fig:EWCSwitht0}. As mentioned in the main text, the key reason for the appearance of such a region is the zero-thickness-shell approximation we adopted, which is in principle required by the holographic thermalization itself. Figure \ref{fig:threestage} shows that the evolution of EWCS can be divided into three stages. The
numerical deformation region occurs in the second stage, which is our focus in this section.

For the second stage, the shell position is between $z_{*}^{b}$ and $z_{*}^{u}$, i.e., if we use $z_{v}$ to label the shell position, then $z_{*}^{b}<z_{v}<z_{*}^{u}$. In this case, we have $v_{b}>0$ and $v_{u}<0$. The action \eqref{eopaction} could be rewritten as follows:
\begin{eqnarray}
\tilde{S}^{dy}_{eop}&=& \mathcal{S}_{1}+  \mathcal{S}_{2}+ \mathcal{S}_{3}\\ \nonumber
&=&\int^{\frac{v_0}{2}}_{v_{b}}\mathcal{S}dv+\int^{-\frac{v_0}{2}}_{\frac{v_0}{2}}\mathcal{S}dv+\int^{v_{u}}_{-\frac{v_0}{2}}\mathcal{S}dv,
\label{eq22}
\end{eqnarray}
where
\begin{equation}
\mathcal{S}=\frac{1}{z^{d-1}L_{AdS}^{d-2}}\left[\sqrt{-f(z,v)-2\frac{dz}{dv}}+2\alpha\left(\frac{dz}{dv}\right)^{2}\left(\sqrt{-f(z,v)-2\frac{dz}{dv}}\right)^{-1}\right].
\label{eq3}
\end{equation}
Note that $v_{b}$ and $v_{u}$ should correspond to the same boundary time. Viewing from the $z-v$ coordinates, this means that we have divided the extreme surface of EWCS into three parts. For the first and third parts, as the integral bounds have same sign so $\partial_{v}f=0$ is guaranteed as the mass function $m(v)$ should be a step function. We only need to show that the second part $\mathcal{S}_{2}$ is limited and can be neglected without causing bad distortions to the calculations. This can easily be understood theoretically as the two phases divided by the zero-thickness shell should be connected at the shell position (i.e., $v=0$), see Figure \ref{fig2}.

\begin{figure}
\centering
 \includegraphics[width=0.5\textwidth]{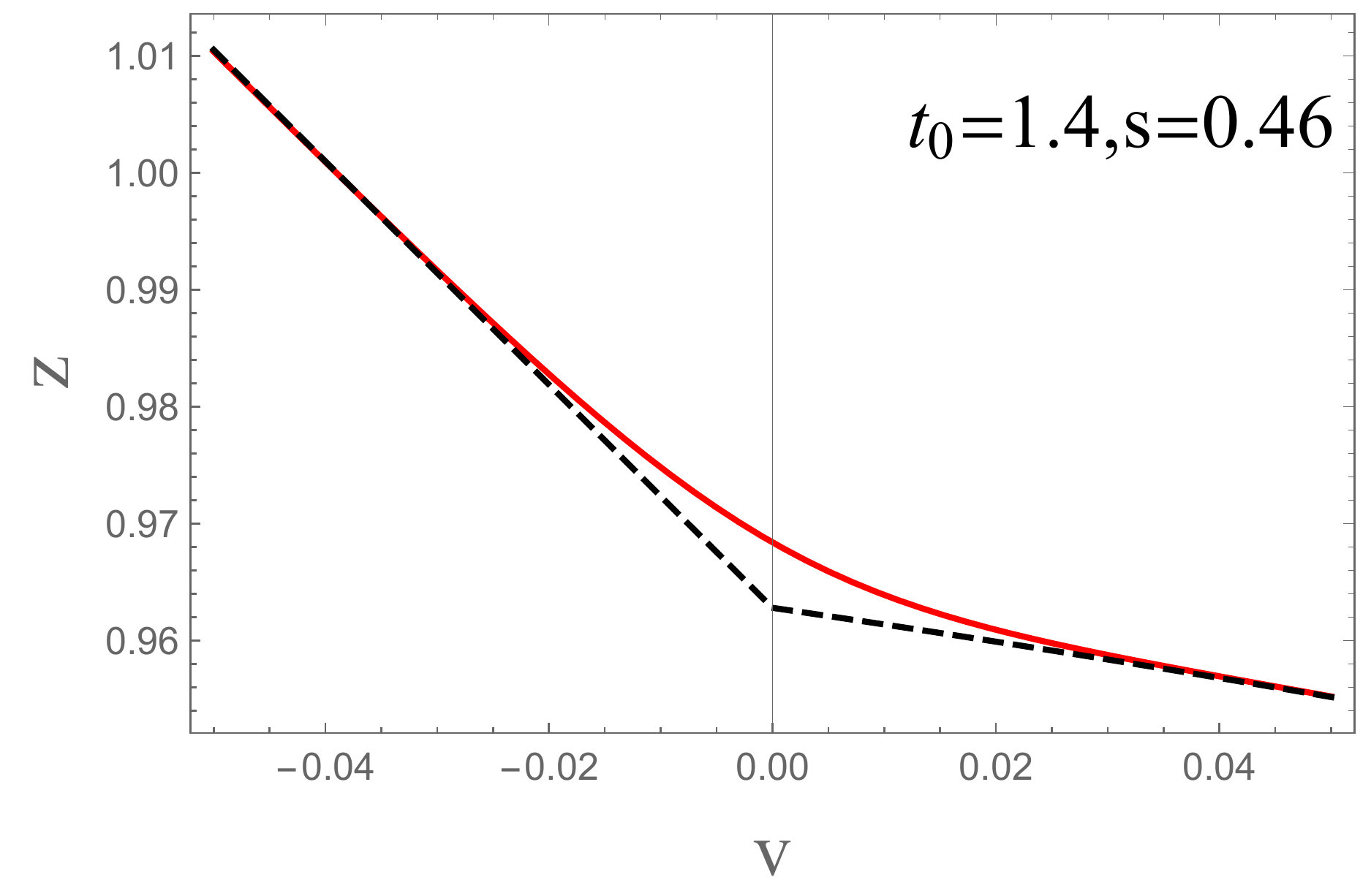}\ \hspace{0.1cm}
	  \caption{A schematic showing the variation for different mass functions around $v=0$. The red line shows the mass function as \eqref{massfunction}. The dashed line represents the mass function $m(v)=H(v)$ where $H(v)$ is a Heaviside unit step function. Here we take $s=0.46$ and $t_0=1.4$.}
	  \label{fig2}
\end{figure}

Thus, we can safely claim that the different mass function or the thickness $v_0$ will not result in undesirable distortions around $v=0$. Therefore, we can rewrite Eqn.~(\ref{eq22}) as follows:
\begin{eqnarray}
\tilde{S}^{dy}_{eop}=\int^{+0}_{v_{b}}\mathcal{S}_{1}dv+\int^{v_{u}}_{-0}\mathcal{S}_{2}dv,
\label{eq4}
\end{eqnarray}
This is what we have adopted in our paper and represents the zero-thickness-shell approximation. Here for $\mathcal{S}_{1}$ we have $f(v,z)=\frac{1}{2\alpha}\left[1-\sqrt{1-4\alpha (1-z^{d})}\right]$, whereas for $\mathcal{S}_{2}$ we have $f(v,z)=\frac{1}{2\alpha}\left[1-\sqrt{1-4\alpha}\right]$, and $z'(v)=-f(z,v)$ for both cases. This is also why we have \eqref{EWt2}.

However, in ordinary way, EWCS should be calculated by directly integrating the action Eqn.~(\ref{eopaction}). Taking $\alpha=-0.05, s=0.46$ and $t_0=1.4$, we plot the $z(v)$ curve in Figure \ref{fig3}. In this initial condition, we have $z^{b}_{*}\simeq 0.543, v^{b}_{*} \simeq 0.842, z^{u}_{*} \simeq 2.402, v_{*}^{u}\simeq -1.330$.
\begin{figure}
\centering
 \includegraphics[width=0.45\textwidth]{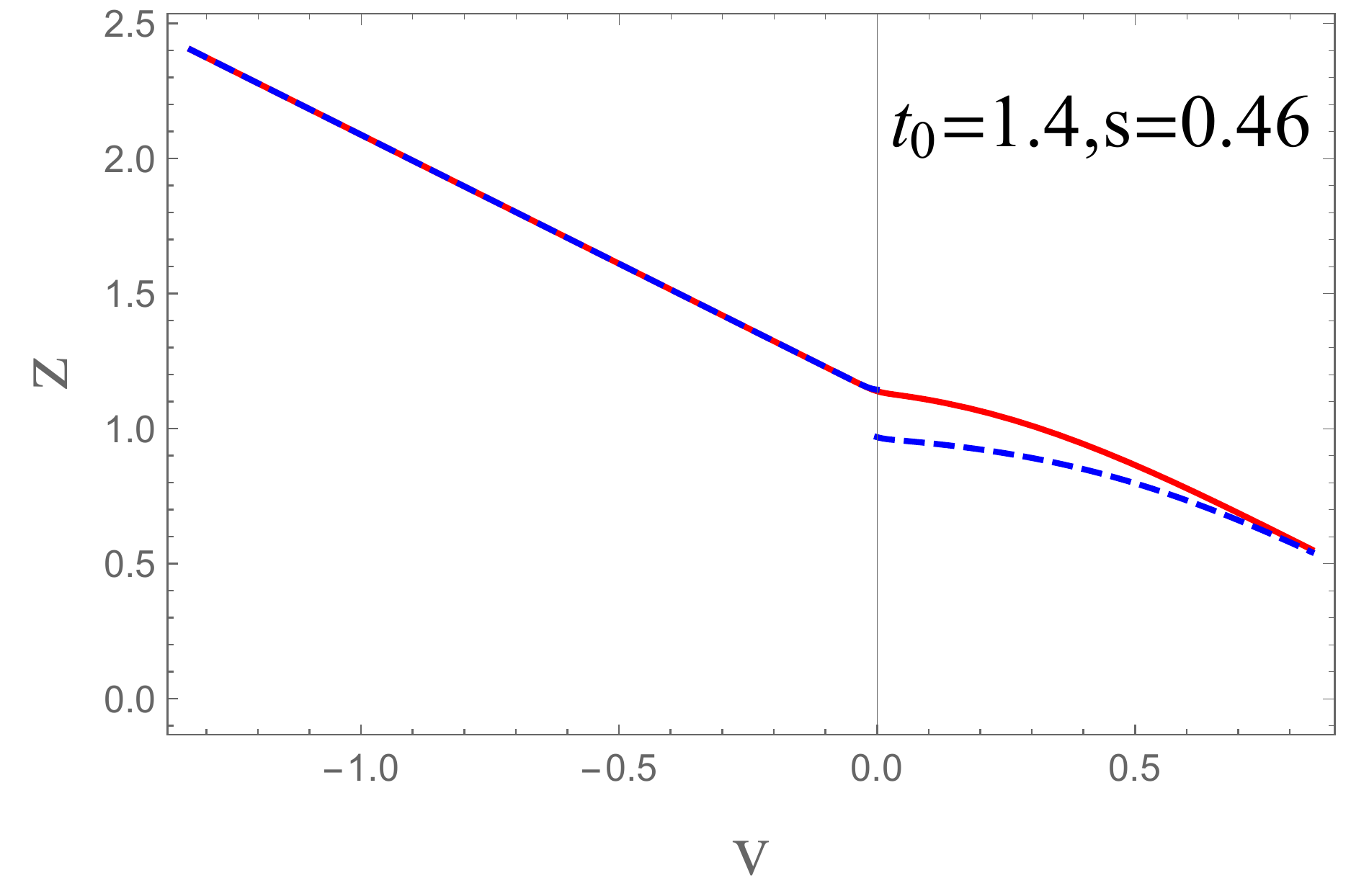}\ \hspace{0.1cm}
 \includegraphics[width=0.45\textwidth]{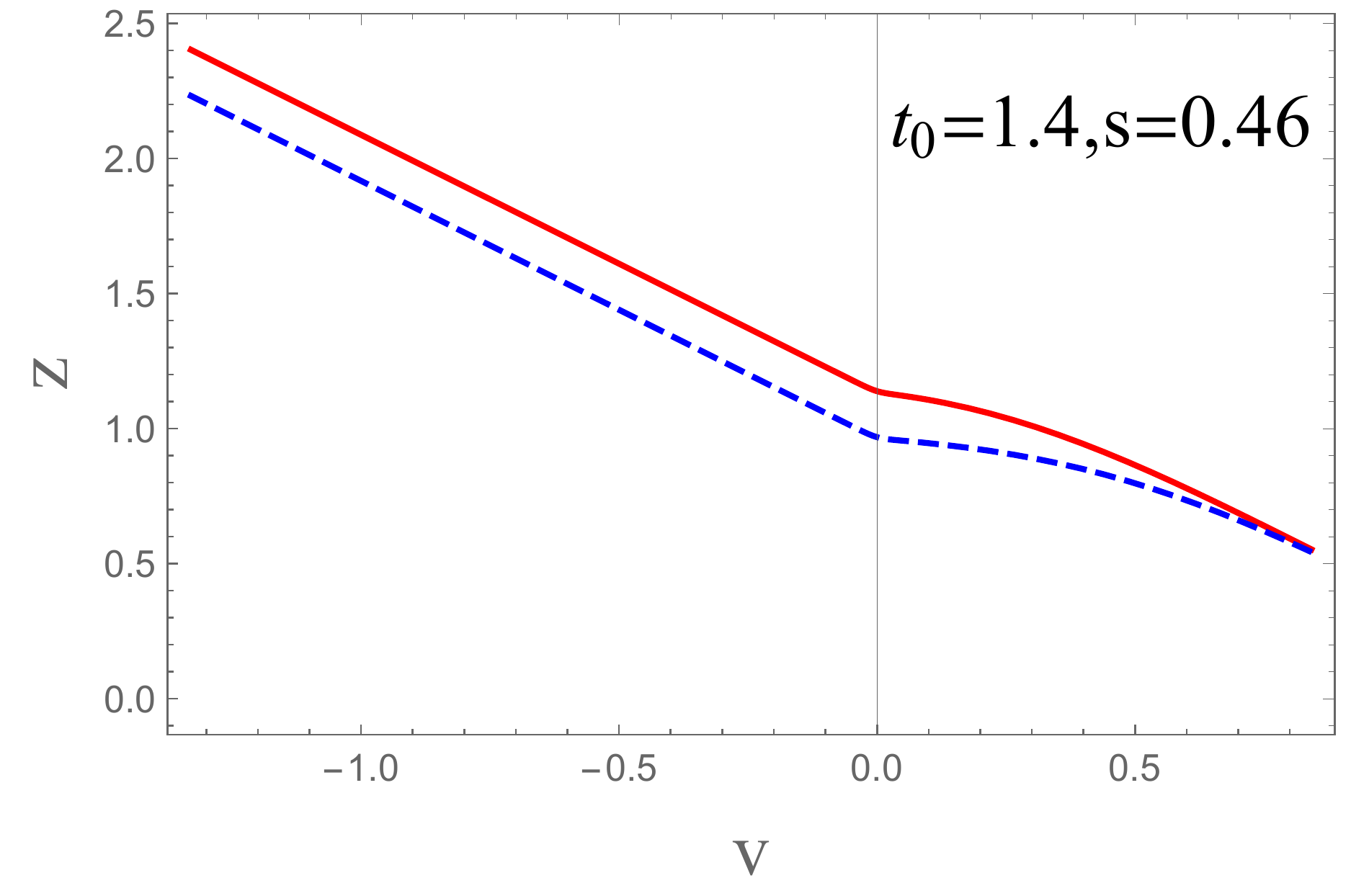}\ \hspace{0.1cm}
	  \caption{$z(v)$ curve. For both figures, the red line is directly calculated from the Euler-Lagrange equation given by the action \eqref{eopaction}. (Left) The dashed line is calculated by choosing the zero-thickness-shell approximation with the action Eqn.~(\ref{eq4}). (Right) The dashed line is calculated by integrating $-f(z,v)=z'(v)$.}
	  \label{fig3}
\end{figure}
For both figures, the red line is directly calculated from the Euler-Lagrange equation given by the action \eqref{eopaction}, by selecting the mass function \eqref{massfunction}.
In the left panel, we approach this curve by using the zero-thickness-shell approximation and Eqn.~(\ref{eq4}), i.e., we separately integrate $-f(z,v)=z'(v)$ according to the integral region
\begin{itemize}
\item{If $v\in{[+0,v_{b}]}$, $z'(v)=\frac{1}{2\alpha}\left[1-\sqrt{1-4\alpha}\right]$, initial condition $z^{b}_{*}\simeq 0.543, v^{b}_{*} \simeq 0.842$,}
\item{If $v\in{[v_{u},-0]}$, $z'(v)=\frac{1}{2\alpha}\left[1-\sqrt{1-4\alpha(1-z^{d})}\right]$, initial condition $z^{u}_{*} \simeq 2.402, v_{*}^{u}\simeq -1.330$.}
\end{itemize}
This guarantees the condition $\partial_{v}f=0$. In the right panel, we approach the red curve by using the solution $-\frac{1}{2\alpha}\left(1-\sqrt{1-4\alpha(1-m(v) z^{4})}\right)=z'(v)$ with the integral region $v\in[v_{u},v_{b}]$.

The difference between two dashed lines can be seen clearly from Figure~\ref{fig3}. A ``jump'' occurred at point $v=0$ in the left panel. By the conditions we have chosen, one obtains $z(0-)\simeq1.133, z(0+)\simeq0.963$. This behavior is consistent with {that in} the right panel of Figure \ref{fig:zvevolution}, where  $z_{v}\approx1.13$ at $t_0=1.4$. Notably, such a jump results from numerically {calculating} HEE with a finite $v_0$, and it may be eliminated if a more accurate numerical method is employed to calculate HEE with $v_0\rightarrow 0$.
While there is no ``jump'' at $v=0$ in the right panel, but unfortunately, the mass function here with a finite $v_0$ is not an analytical solution for the functional \eqref{eopaction}. Here, one can find that $z(0^+)=z(0^-)<1$, so in this case there will be no numerical deformation region.

The extremal surface of EWCS in $z-v$ should be analytically and physically given by the functional \eqref{eopaction} with a zero-thickness shell. Such an extremal surface, given by $-\frac{1}{2\alpha}\left(1-\sqrt{1-4\alpha(1-H(v) z^{4})}\right)=z'(v)$, should behave like the dashed line in the right panel of Figure \ref{fig3}.

In summary, the numerical deformation region appears due to the zero-thickness-shell approximation when we try to analytically calculate the area of EWCS. Further, it will be interesting to see if such a region persists in other boundary shapes of the boundary regions on the EWCS evolution, which we will leave this to future research.
%%%%%%%%%%%%%%%%%%%%%%%%%%%%%%%%%%%%%%%%%%%%%%%%%%%%

%\bibliography{GB_draft_300820}

\end{document}